\documentclass[preprint,showkeys,amsmath,longbibliography]{revtex4-1}
\usepackage{morefloats}
\usepackage{psfrag}
\usepackage{graphicx}
\usepackage{subfigure}
\usepackage{array}
\usepackage{natbib} 
\def\be{\begin{equation}}       \def\ee{\end{equation}}
\def\bea{\begin{eqnarray}}      \def\eea{\end{eqnarray}}

\begin{document}
\title{
Manipulation of two quantum systems through their mutual interaction in presence of a radiation field}
\author{Gehad Sadiek\footnote{Corresponding author: gsadiek@sharjah.ac.ae}}
\affiliation{Department of Applied Physics and Astronomy, University of Sharjah, Sharjah 27272, UAE and\\
Department of Physics, Ain Shams University, Cairo 11566, Egypt}
\author{Wiam AlDrees}
\affiliation{Department of Physics and Astronomy, King Saud University, Riyadh 11451, Saudi Arabia} 
\author{M. Sebaweh Abdallah\footnote{Deceased.}}
\affiliation{Department of Mathematics, King Saud University, Riyadh 11451, Saudi Arabia}
\begin{abstract}
In cavity QED, the mutual interaction between natural atomic systems in presence of a radiation field was ignored due to its negligible impact compared with the coupling to the field. The newly engineered artificial atomic systems (such as quantum dots and superconducting circuits) proposed for quantum information processing enjoy strong interaction with same type of systems or even with other types in hybrid structures, which is coherently controllable and moreover they can be efficiently coupled to radiation fields. We present an exact analytic solution for the time evolution of a composite system of two interacting two-level quantum systems coupled to a single mode radiation field, which can be realized in cavity (circuit) QED. We show how the non-classical dynamical properties of the composite system are affected and can be tuned by introducing and varying the mutual coupling between the two systems. Particularly, the collapse-revival pattern shows a splitting during the revival intervals as the coupling ratio (system-system to system-field) increases, which is a sign of an interruption in the system-radiation energy exchange process. Furthermore, the time evolution of the bipartite entanglement between the two systems is found to vary significantly depending on the coupling ratio as well as the initial state of the composite system resulting in either an oscillatory behavior or a collapse-revival like pattern. Increasing the coupling ratio enhances the entanglement, raises its oscillation average value and emphasizes the collapse-revival like pattern. However, raising the coupling ratio beyond unity increases the revival time considerably. The effect of the other system parameters such as detuning and radiation field intensity on the system dynamics has been investigated as well.
\end{abstract}
\keywords{Atomic and Molecular Physics, photonics, Quantum information}
\maketitle
\section{Introduction}
\label{Introduction}
The interaction between a quantum system and a bosonic field has been one of the central problems in physics. It manifests itself in many different systems of interest such as spin coupling to phonon modes in crystalline lattices \cite{Holstein1959_1, Holstein1959_2}, the interaction between atoms, ions or molecules with radiation field in cavity QED \cite{Scully-Zubairy1997_B, Blatt2008}, semiconductor quantum dots interacting with radiation field in cavity QED and optical nanocavities \cite{Hanson2008, Gywat2010, Laucht2010, Lodahl2015}, superconducting Josephson junction qubits in QED circuits \cite{Paik2011} and the manipulation of the cold atoms (ions or molecules) using microwave radiation field~\cite{Hood2000, Hanson2008, Soderberg2009, Liu2016}. Most of these systems have been proposed as very promising candidates for the future underlying technology of quantum information processing and quantum computations \cite{Nielsen-Chuang2010_B}. For any of these systems to play this role it has to be prepared as a two-state quantum system (qubit), addressed and coherently coupled to the other systems in a controllable manner (using a gate voltage or a radiation field for instance).

One of the most successful approaches in describing the interaction between a two-level quantum system and a bosonic field is the Rabi model \cite{Rabi1936}. As the Rabi model itself is not analytically solvable, different treatments and approximations were introduced to simplify it. The Jaynes-cummings model (JCM)~\cite{Jaynes1963}, which is exactly integrable, was derived from the Rabi model using the rotated wave approximation (RWA)~\cite{Irish2007} and has been widely used to describe the interaction between quantum systems and radiation fields. The original JCM was concerned with the interaction between a single two-level quantum system (atom) and a single-mode quantized radiation field. The validity of the JCM is guaranteed as long as the coupling between the quantum system and the radiation field is much smaller than the field frequency and the system energy gap (weak and strong coupling regimes) but it fails when they become of the same order of magnitude (ultra-strong coupling regime). The JCM exhibits a number of interesting nonclassical effects, such as the collapse-revival phenomenon~\cite{Eberly1980,Yoo1985,Rempe1987}, sub-Poissonian photon statistics~\cite{Kim1993}, atom-field entanglement~\cite{Kudryavtsev1993}, and squeezing~\cite{Meystre1982}. The JCM was extended to include more than one atom as well as multi-level atom coupled  to a radiation field by Tavis and Cummings~\cite{Tavis1968} (TJCM model) and implemented by many authors latter \cite{Dicke1954, Cummings1983,Mahmood1987,Jex1992}. Particularly, the case of two two-level atoms interacting with a single-mode radiation field was studied intensively and found to exhibit a more complicated physical properties compared to the one-atom case (JCM) in the collapse-revival phenomenon\cite{Iqbal1988, Deng1985, Prakash2008}, atom-atom entanglement~\cite{Zheng2000,Osnaghi2001} and atomic squeezing~\cite{Prakash2007, Orany2008,Hekmatara2014,Liao2011}. The JCM model (and the subsequent TJCM model) has been always the main mathematical frame to study the interaction between atoms and radiation fields in cavity QED. The interaction among the natural atoms themselves within the cavity was always ignored as it was considered as extremely week compared with the coupling between each atom and the radiation field.

The great interest in realizing quantum information processing systems in the last few decades sparked intense efforts and led to a significant progress in engineering new quantum systems that are considered as very promising candidates for playing the role of a qubit. These developed systems (such as semiconducting quantum dots and superconducting circuits) in addition to some natural atomic systems (such as Rydberg atoms and trapped atoms, ions and molecules), in contrary to the natural conventional atoms, enjoy a strong coupling with a similar type of system or even with a different type (when implemented in a hybrid system) \cite{Buluta2011, Xiang2013, Lodahl2015} through direct or mediated interaction. Interestingly, even the natural atoms were forced to interact with each other by preparing them in arrays of ultra cold paired atoms in an optical lattice and induce controlled exchange coupling between each pair by placing the two atoms into the same physical location \cite{Anderlini2007} and even the super-exchange coupling between the atoms was controlled in sign and magnitude \cite{Trotzky2008}. All these systems can be coupled to electromagnetic radiations and exchange energy with them in QED cavity. Cavity QED is concerned with the interaction between an electromagnetic radiation and a quantum system where the quantization of the radiation field is crucial \cite{Scully-Zubairy1997_B, Dutra2005}. As a result, the cavity can serve as a data bus in a quantum information processing systems where it carries information between the different involved quantum systems (qubits). Practically, only natural atomic systems and spins in quantum dots can be coupled to traditional optical cavity \cite{Hood2001, Buck2003, Ozdemir2011} whereas the superconductor systems couple to, the cavity equivalent, SC resonators (SC coplanar waveguide (CPW) \cite{Wallraff2004,Hofheinz2009} and $LC$ resonators \cite{Zhou2008,Liao2010}). The coupling between each one of these systems and the radiation field can be described using the TJCM Hamiltonian, whereas the interaction between these quantum systems themselves (either like or unlike systems) can be effectively described by the spin-1/2 Heisenberg exchange Hamiltonian for either direct or mediated coupling \cite{Xiang2013, Blatt2012, Blatt2008, Lodahl2015}.

In this paper, we consider two identical mutually interacting two-level quantum systems coupled to a single-mode quantized radiation field. We present an exact analytic solution for the time evolution of the system. We investigate how the presence of interaction between the two quantum systems may influence the well known and studied dynamical properties of the system such as the collapse-revival phenomenon in population inversion and entanglement.
We show how the collapse-revival pattern of any of the two quantum systems is affected by introducing and varying the interaction strength between the two quantum systems, where the revival oscillations split into smaller ones as the coupling strength is increased which is a sign of an interruption in the energy exchange process between the radiation field and the quantum systems due to the interaction. Increasing the radiation intensity causes a longer collapse time and makes the collapse-revival pattern more robust to the system-system interaction effect. Furthermore, we investigate the effect of the interplay between the two couplings (system-system and system-radiation) on the bipartite entanglement between the two quantum systems. We demonstrate how increasing the coupling ratio (system-system to system-radiation) may enhance the entanglement and affects its dynamical behavior, which shows either regular oscillatory or collapse-revival like pattern based on the initial state of the system. Also, we explore and compare the effects of zero and non-zero detuning on the system dynamical properties.

This paper is organized as follows. In Sec.II, we discuss our model. In Sec. III, we present an exact analytic solution for the time evolution of the system. We implement our solution to study the collapse-revival phenomenon in the population inversion and the bipartite entanglement dynamics in Secs. IV, and V respectively. We conclude in Sec. VI. 
\section{The Model}
\label{The Model}
\begin{figure}[ht]
\centering
\includegraphics[width=13cm]{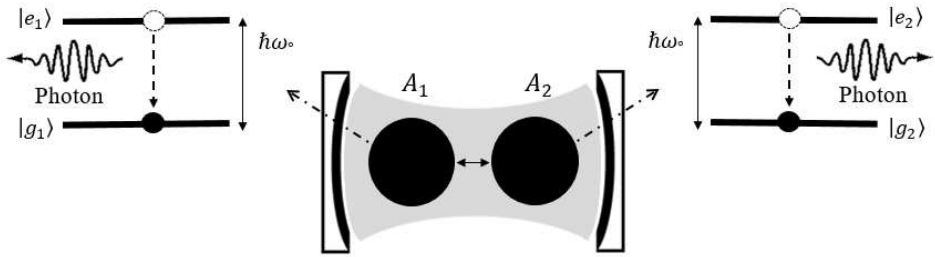}
\caption{{\protect\footnotesize Two two-level coupled quantum systems in cavity QED (or equivalently circuit QED).}}
\label{Fig1ch3}
\end{figure}

We consider a system of two identical quantum systems, illustrated in Fig.~\ref{Fig1ch3}, each one of them is characterized by two levels: ground $\left|g_{i}\right\rangle$ and excited $\left|e_{i}\right\rangle$, where $i=1,2$ corresponds to the first and second system respectively. The two systems are coupled to the same single-mode quantized radiation field with the same coupling constant $\lambda_1$. The coupling between the two quantum systems is modeled as an isotropic $XY$ exchange interaction between two spin-1/2 particles with coupling strength $\lambda_2$. The Hamiltonian of the composite system is given by
\begin{equation}
\hat{H} = \hat{H}_{field} + \hat{H}_{system} + \hat{H}_{system-field} + \hat{H}_{system-system} \; ,  
\label{eq:H}
\end{equation}
where
\begin{eqnarray}
\nonumber &\hat{H}&_{field} = \omega \; \hat{a}^\dagger \hat{a} \;, \\
\nonumber &\hat{H}&_{system}  = \frac{\omega_{\circ}}{2} \sum_{i=1,2} \; \hat{\sigma}^{(i)}_{z}\;, \\
\nonumber &\hat{H}&_{system-field} =  \lambda_1 \sum_{i=1,2} \; (\hat{a}\hat{\sigma}^{(i)}_{+}+\hat{a}^\dagger \hat{\sigma}^{(i)}_{-}) \;, \\
&\hat{H}&_{system-system} = \lambda_2 \; (\hat{\sigma}^{(1)}_{-}\hat{\sigma}^{(2)}_{+}+\hat{\sigma}^{(1)}_{+}\hat{\sigma}^{(2)}_{-})\; ,
\label{eq:H_terms}
\end{eqnarray}
The first and second terms in the Hamiltonian represent the free quantized radiation field and the non-interacting two quantum systems while the third and fourth terms represent the system-field and system-system interactions respectively. $\omega$ and $\omega_{\circ}$ are the frequencies of the single-mode radiation field and the quantum system transition respectively, $\hat{a}^\dagger$ and $\hat{a}$ are creation and annihilation operators of the radiation field which satisfy the commutation relation $[\hat{a},\hat{a}^\dagger]=1$ and $\hat{\sigma}^{(i)}_{\pm}$ and $\hat{\sigma}^{i}_{z}$ are the usual Pauli spin operators of the $i$th quantum system.
\\
Using the Heisenberg equation of motion, which for any operator $\hat{O}$ reads
\begin{equation}
\frac{d\hat{O}}{dt}=\textrm{-i}[\hat{Q},\hat{H}] + \frac{\partial \hat{O}}{\partial t},
\label{eq:Heisenberg_eqn_of motion}
\end{equation}
one obtains the following equations for the field and system operators
\begin{eqnarray}
\nonumber \frac{d\hat{a}}{dt} &=& -\textrm{i} \omega \hat{a} - \textrm{i} \lambda_{1} (\hat{\sigma}^{(1)}_{-}+\hat{\sigma}^{(2)}_{-}),\\
\nonumber \frac{d\hat{\sigma}^{(1)}_{-}}{dt} &=& -\textrm{i} \omega_{\circ} \hat{\sigma}^{(1)}_{-} + \textrm{i} \lambda_{1} \hat{a} \hat{\sigma}^{(1)}_{z} +  \textrm{i} \lambda_{2} \hat{\sigma}^{(1)}_{z} \hat{\sigma}^{(2)}_{-},\\
\nonumber \frac{d\hat{\sigma}^{(2)}_{-}}{dt} &=& -\textrm{i} \omega_{\circ} \hat{\sigma}^{(2)}_{-} + \textrm{i} \lambda_{1} \hat{a} \hat{\sigma}^{(2)}_{z} +  \textrm{i} \lambda_{2} \hat{\sigma}^{(1)}_{-}\hat{\sigma}^{(2)}_{z},\\
\nonumber \frac{d\hat{\sigma}^{(1)}_{z}}{dt} &=& 2\textrm{i} \lambda_{1} (\hat{a}^\dagger \hat{\sigma}^{(1)}_{-}-\hat{a}\hat{\sigma}^{(1)}_{+})+ 2\textrm{i} \lambda_{2} (\hat{\sigma}^{(1)}_{-}\hat{\sigma}^{(2)}_{+}-\hat{\sigma}^{(1)}_{+}\hat{\sigma}^{(2)}_{-}),\\
\frac{d\hat{\sigma}^{(2)}_{z}}{dt} &=& 2\textrm{i} \lambda_{1} (\hat{a}^\dagger \hat{\sigma}^{(2)}_{-}-\hat{a}\hat{\sigma}^{(2)}_{+})+ 2\textrm{i} \lambda_{2} (\hat{\sigma}^{(1)}_{+}\hat{\sigma}^{(2)}_{-}-\hat{\sigma}^{(1)}_{-}\hat{\sigma}^{(2)}_{+}),
\label{eq:opertors_dynamics}
\end{eqnarray}
Assuming that initially the quantum systems are in a pure state and the field is in the coherent state, the wave function of the composite system at $t=0$ can be written as
\begin{equation}
\vert\psi(0)\rangle = [a\; \vert e_{1},e_{2}\rangle + b \; \vert e_{1},g_{2}\rangle + c \; \vert g_{1},e_{2}\rangle + d \; \vert g_{1},g_{2}\rangle]\otimes \vert\alpha\rangle,
\label{eq:8}
\end{equation}
where $a,b,c$ and $d$, are arbitrary complex quantities that satisfy the condition 
\begin{equation}
\vert a \vert^2 + \vert b \vert^2 +\vert c \vert^2 +  \vert d \vert^2 =1,
\label{eq:9}
\end{equation}
and $\vert\alpha\rangle$ is the coherent state defined as
\begin{equation}
\vert\alpha\rangle=\sum_{n} Q_{n} \vert n \rangle;   \qquad Q_{n}=\frac{\alpha^{n}}{\sqrt{ n !}} \exp\left(-\frac{\vert \alpha \vert^2}{2}\right),
\label{eq:10}
\end{equation}
where $\vert \alpha \vert^2=\overline{n}$ is the mean photon number and $\vert n \rangle$ are the photon number states, which satisfy the relations: $\hat{a}^{\dagger} \vert n \rangle =\sqrt{n+1}\vert n+1 \rangle$ and $\hat{a} \vert n+1 \rangle = \sqrt{n+1} \vert n \rangle$. The wave function at any time $t$ latter can be written as
\begin{eqnarray}
\nonumber \vert\psi(t)\rangle &=& \sum_{n} [ A_{n}(t)\vert e_{1},e_{2},n\rangle + B_{n+1}(t)\vert e_{1},g_{2},n+1\rangle + C_{n+1}(t) \vert g_{1},e_{2},n+1\rangle \\
&& + D_{n+2}(t)\vert g_{1},g_{2},n+2\rangle ],
\label{eq:psi_t}
\end{eqnarray}
where $ \vert e_{1},e_{2},n\rangle $ is the state in which the two quantum systems are in excited state and the field has $n$ photons, $ \vert e_{1},g_{2},n+1\rangle $ is the state in which the first one is in the excited state and the second is in the ground state and the field has $n+1$ photons and so on. The quantum system sates satisfy the relations $\hat{\sigma}_{+} \vert g \rangle = \vert e \rangle$, $\hat{\sigma}_{-} \vert g \rangle =0$, $\hat{\sigma}_{+} \vert e \rangle = 0$ and $\hat{\sigma}_{-} \vert e \rangle =  \vert g \rangle$.
The time-dependent coefficients $A_{n}(t), B_{n+1}(t), C_{n+1}(t)$ and $D_{n+2}(t)$ can be obtained by solving the Schr{\"o}dinger equation of the composite system, which will be discussed in the next section. 

Once we obtain the system wave function $\vert\psi(t)\rangle$, we can calculate the composite system density matrix $\hat{\rho}(t)=\vert\psi(t)\rangle\langle\psi(t)\vert$. The reduced density matrix of the two quantum systems, $\hat{\rho}_{\textrm{qs}}(t)$, can be obtained by tracing out the field 
\begin{equation}
\hat{\rho}_{\textrm{qs}}(t)=\textrm{Tr}_{\textrm{field}}\; \hat{\rho}(t)= \sum_{l} \langle l \vert \psi(t)\rangle \langle \psi(t) \vert l \rangle.
\label{eq:qs_rdm}
\end{equation}
\section{The Analytic Solution}
\label{The Analytic Solution}
We devote this section to solve the Schr{\"o}dinger equation of the system and provide an exact analytic expression for the time-dependent coefficients of the system wave function. We start by rewriting the Hamiltonian (Eq.~(\ref{eq:H})) as 
\begin{equation}
\hat{H}=\hat{H}_{\circ}+\hat{H}_{int}\;,
\label{eq:14}
\end{equation}
where 
\begin{equation}
\hat{H}_{\circ} = \omega \; \hat{N} + \frac{\Delta}{2} \sum_{i=1,2} \; \hat{\sigma}^{(i)}_{z}\;,
\label{eq:15}
\end{equation} 
\begin{equation}
\hat{H}_{int}= \lambda_1 \sum_{i=1,2} \; (\hat{a}\hat{\sigma}^{(i)}_{+}+\hat{a}^\dagger \hat{\sigma}^{(i)}_{-})\: + \lambda_2 \; (\hat{\sigma}^{(1)}_{-}\hat{\sigma}^{(2)}_{+}+\hat{\sigma}^{(1)}_{+}\hat{\sigma}^{(2)}_{-}),
\label{eq:16}
\end{equation}
and
\begin{equation}
\hat{N}= \hat{a}^\dagger \hat{a} + \frac{1}{2} \sum_{i=1,2} \; \hat{\sigma}^{(i)}_{z}\;, 
\label{eq:5}
\end{equation}
where $\hat{N}$ is a constant of motion and $\Delta=\omega_{\circ}-\omega$ is the detuning parameter.

It is more convenient to work in the interaction picture where we define $\hat{V}_I=\hat{U} \hat{H}_{int} \hat{U}^{\dagger}$ with $\hat{U}=e^{\textrm{i}\hat{H}_{\circ}t}$. As a result, we obtain
\begin{eqnarray}
\nonumber \hat{V}_{I}(t)= \; \lambda_1 \sum_{i=1,2}\; &(&\hat{a}\;e^{\textrm{i}\Delta t} \hat{\sigma}^{(i)}_{+} + \hat{a}^\dagger \; e^{-\textrm{i}\Delta t} \hat{\sigma}^{(i)}_{-})+ \hbar \: \lambda_2 ( \hat{\sigma}^{(1)}_{-}\hat{\sigma}^{(2)}_{+}\\
&&+ \hat{\sigma}^{(1)}_{+}\hat{\sigma}^{(2)}_{-})\;.
\label{eq:V_int}
\end{eqnarray}
Now, substituting $\vert\psi(t)\rangle$ and $V_I(t)$ into Schr{\"o}dinger equation
\begin{equation}
\textrm{i} \; \frac{\partial}{\partial t}\vert\psi(t)\rangle = \hat{V}_I(t)\vert\psi(t)\rangle, 
\label{eq:Sch_eqn}
\end{equation}
yields a system of coupled differential equations
\begin{eqnarray}
\nonumber \textrm{i}\dot{A}_{n}(t) &=& \alpha\; e^{\textrm{i}\Delta t} \;(B_{n+1}(t) + C_{n+1}(t)),\\
\nonumber \textrm{i}\dot{B}_{n+1}(t) &=& \alpha \;e^{-\textrm{i} \Delta t }\;A_{n}(t) + \beta \;e^{\textrm{i}\Delta t}\;D_{n+2}(t) + \lambda_{2}\;C_{n+1}(t),\\
\nonumber \textrm{i}\dot{C}_{n+1}(t) &=& \alpha\;e^{-\textrm{i} \Delta t }\;A_{n}(t) + \beta \;e^{\textrm{i}\Delta t}\;D_{n+2}(t) + \lambda_{2}\;B_{n+1}(t),\\
\textrm{i}\dot{D}_{n+2}(t)&=& \beta\; e^{-\textrm{i}\Delta t}\;(B_{n+1}(t) + C_{n+1}(t)),
\label{eq:DE_sys1}
\end{eqnarray}
where $\alpha=\lambda_{1}\sqrt{n+1}$ and $\beta=\lambda_{1}\sqrt{n+2}$. Substituting $ K(t)= B_{n+1}(t)+C_{n+1}(t)$, Eqs.~(\ref{eq:DE_sys1}) simplify to
\begin{eqnarray}
\nonumber \textrm{i}\dot{A}_{n}(t) &=& \alpha K(t)\;e^{\textrm{i}\Delta t},\\
\nonumber \textrm{i}\dot{D}_{n+2}(t)&=& \beta K(t)\;e^{-\textrm{i}\Delta t},\\
\textrm{i}\dot K(t) &=& 2 \alpha\;e^{-\textrm{i} \Delta t}\;A_{n}(t) + 2\beta e^{\textrm{i}\Delta t} D_{n+2}(t) + \lambda_{2} K(t),
\label{eq:DE_sys2}
\end{eqnarray}
which, after some calculations becomes
\begin{eqnarray}
\nonumber \dddot K(t) &+& \textrm{i}\lambda_{2}\ddot K(t) + [2(\alpha^{2}+\beta^{2})+\Delta^{2}]\;\dot K(t) - \textrm{i}[2\Delta(\alpha^{2}-\beta^{2})\\
&&-\lambda_{2}\Delta^{2}]\;K(t)=0,
\label{eq:3rd_order_DE}
\end{eqnarray}
with a solution
\begin{equation}
K(t)= \sum^{3}_{j=1}\delta_{j} e^{m_{j}t},
\label{eq:23}
\end{equation}
where
\begin{eqnarray}
\nonumber \delta_{1}&=&(B_{n+1}(0)+C_{n+1}(0))-(\delta_{2}+\delta_{3}),\\
\nonumber \delta_{2}&=&\frac{1}{(m_{1}-m_{2})(m_{3}-m_{2})} \lbrace 2\alpha A_{n}(0)[\textrm{i}(m_{1}+m_{3})-\lambda_{2}-\Delta]+2\beta D_{n+2}(0) [\textrm{i}(m_{1}+m_{3})\\
\nonumber &&- \lambda_{2}+\Delta]+ [\textrm{i}(m_{1}+m_{3})(\lambda_{2}-\textrm{i}m_{1})-2(\alpha^{2}+\beta^{2})-\lambda^{2}_{2}-m_{1}^{2}]\\
\nonumber && \times(B_{n+1}(0)+C_{n+1}(0))\rbrace, \\
\nonumber \delta_{3}&=&\frac{1}{(m_{1}-m_{3})(m_{2}- m_{3})}\lbrace 2\alpha A_{n}(0)[\textrm{i}(m_{1}+m_{2})-\lambda_{2}-\Delta]+2\beta D_{n+2}(0)[\textrm{i}(m_{1}+m_{2})\\
\nonumber &&-\lambda_{2}+\Delta]+[\textrm{i}(m_{1}+m_{2})(\lambda_{2}-\textrm{i}m_{1})-2(\alpha^{2}+\beta^{2})-\lambda_{2}^{2}-m_{1}^{2}]\\
&& \times(B_{n+1}(0)+C_{n+1}(0))\rbrace,
\label{eq:24}
\end{eqnarray}
and 
\begin{eqnarray}
\nonumber m_{1}&=&(v_{1}+v_{2})-\textrm{i}\frac{\lambda_{2}}{3},\\
\nonumber m_{2}&=&-\frac{v_{1}+v_{2}}{2}+\textrm{i} \frac{\sqrt{3}}{2}(v_{1}-v_{2})-\textrm{i}\frac{\lambda_{2}}{3},\\
m_{3}&=&-\frac{v_{1}+v_{2}}{2}-\textrm{i} \frac{\sqrt{3}}{2}(v_{1}-v_{2})-\textrm{i}\frac{\lambda_{2}}{3},
\label{eq:ms}
\end{eqnarray}
where
\begin{equation}
v_{1}=[ -\frac{\mu}{2}+(\frac{\mu^{2}}{4}+\frac{\eta^{3}}{27})^\frac{1}{2}]^\frac{1}{3}; \qquad
v_{2}=[-\frac{\mu}{2}-(\frac{\mu^{2}}{4}+\frac{\eta^{3}}{27})^\frac{1}{2}]^\frac{1}{3},
\label{eq:v1_v2}
\end{equation}
and 
\begin{equation}
\mu=-\frac{\textrm{i}}{27} [2\lambda_{2}^{3}+18\lambda_{2}(\alpha^{2}+\beta^{2}-\Delta^2)+54 \Delta (\alpha^{2}-\beta^{2})],
\label{eq:mu}
\end{equation}
\begin{equation}
\eta=\frac{1}{3}[6(\alpha^{2}+\beta^{2})+3\Delta^2+\lambda_{2}^{2}]\;.
\label{eq:eta}
\end{equation}
Finally, the solution of the set of differential equations~(\ref{eq:DE_sys1}) takes the form
\begin{eqnarray}
\nonumber A_{n}(t) &=& A_{n}(0)-\textrm{i}\alpha \sum^{3}_{j=1}[\frac{\delta_{j}}{m_{j}+\textrm{i}\Delta}( e^{(m_{j}+\textrm{i}\Delta )t}-1)],\\
\nonumber B_{n+1}(t)&=&\frac{1}{2}[(B_{n+1}(0)-C_{n+1}(0))e^{\textrm{i}\lambda_{2}t}+\sum^{3}_{j=1}\delta_{j} e^{m_{j}t}],\\
\nonumber C_{n+1}(t)&=&\frac{1}{2}[(C_{n+1}(0)-B_{n+1}(0))e^{\textrm{i}\lambda_{2}t}+\sum^{3}_{j=1}\delta_{j} e^{m_{j}t}],\\
D_{n+2}(t)&=& D_{n+2}(0)-\textrm{i}\beta \sum^{3}_{j=1}[\frac{\delta_{j}}{m_{j}-\textrm{i}\Delta}( e^{(m_{j}-\textrm{i}\Delta )t}-1)],
\label{eq:coef_soln}
\end{eqnarray}
where the initial values of the coefficients are given by
\begin{eqnarray}
A_{n}(0)=Q_{n}\;a, \;\;\; B_{n+1}(0)=Q_{n+1}\;b,\;\;\; C_{n+1}(0)=Q_{n+1}\;c,\;\;\; D_{n+2}(0)=Q_{n+2}\;d.
\label{eq:init_values}
\end{eqnarray}

As can be noticed for Eqs.~(\ref{eq:coef_soln}) to represent an acceptable physical solution, the parameters $m_1$, $m_2$ and $m_3$ in the exponents can have only either negative or imaginary values, otherwise the coefficients will blow up with time. This restriction causes certain roots of $v_1$ and $v_2$ in Eqs.~(\ref{eq:v1_v2}) to be appropriate for the solution whereas the others represent a non-physical solution. In fact, each one of the two quantities $v_1$ and $v_2$ will have three, generally complex, roots. Therefore $v_{1}$ and $ v_{2}$ defined by Eqs.~(\ref{eq:ms}) has nine possible combinations, only six of them lead to physically acceptable solution. Nevertheless, very fortunately these six combinations enable us to span the whole parameter space of the system.

Finally the reduced density matrix of the two quantum systems defined by Eq.~(\ref{eq:qs_rdm}) can be obtained, utilizing that $\rho^{\dagger}=\rho$, as
\begin{equation}
\rho_{qs}= \displaystyle \sum_{n=0}^{\infty} \left( \begin{matrix}
\vert A_{n} \vert^2 & A_{n+1}B_{n+1}^{*} & A_{n+1}C_{n+1}^{*} & A_{n+2}D_{n+2}^{*}\\
B_{n+1}A_{n+1}^{*} & \vert B_{n+1} \vert^2 & B_{n+1}C_{n+1}^{*} & B_{n+2}D_{n+2}^{*}\\
C_{n+1}A_{n+1}^{*} & C_{n+1}B_{n+1}^{*} & \vert C_{n+1} \vert^2 & C_{n+2}D_{n+2}^{*}\\
D_{n+2}A_{n+2}^{*} & D_{n+2}B_{n+2}^{*} & D_{n+2}C_{n+2}^{*} &\vert D_{n+2} \vert^2
\end{matrix} \right),
\label{qs_rdm}
\end{equation}

\section{Atomic inversion}
\label{Atomic inversion}
The collapse-revival behavior represents one of the most important non-classical phenomena in the field of quantum optics; it is usually observed when atomic inversion, as result of the interaction between the field and the atom within a cavity, is investigated. In the early days of the quantum optics theory, the atom-field interaction was treated semi-classically, where the atom was assumed to be quantized but not the field. This treatment showed that the atom excitation probability will exhibit a continuous oscillation which was called the Rapi oscillation (cycles). A full quantum mechanical treatment of the atom-field interaction assuming the field has discrete states, described by JCM, demonstrated that this oscillation is not really continuous but disappears and rebuilds again in what latter became known as the collapse-revival phenomenon \cite{Scully-Zubairy1997_B}. This oscillation was realized experimentally for the first time in 1986 \cite{Rempe1987}. Atomic inversion is defined as the expectation value of the operator $\hat{\sigma}_{z}$ or the difference between the probabilities of finding the atom in its excited state and ground state.
\begin{figure}[htbp]
\begin{minipage}[c]{\textwidth}
\centering
\subfigure{\includegraphics[width=9cm]{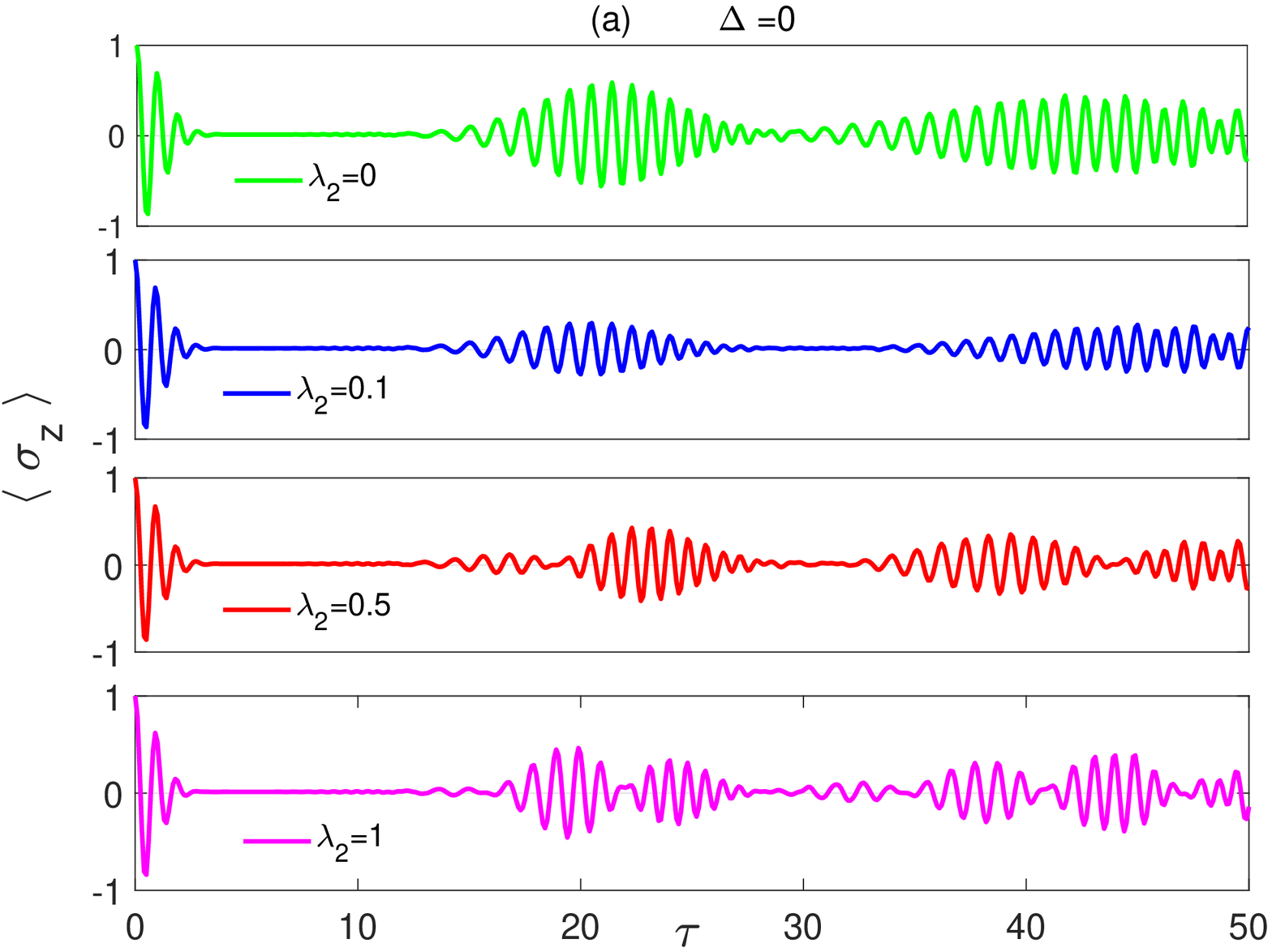}} \quad
\subfigure{\includegraphics[width=9cm]{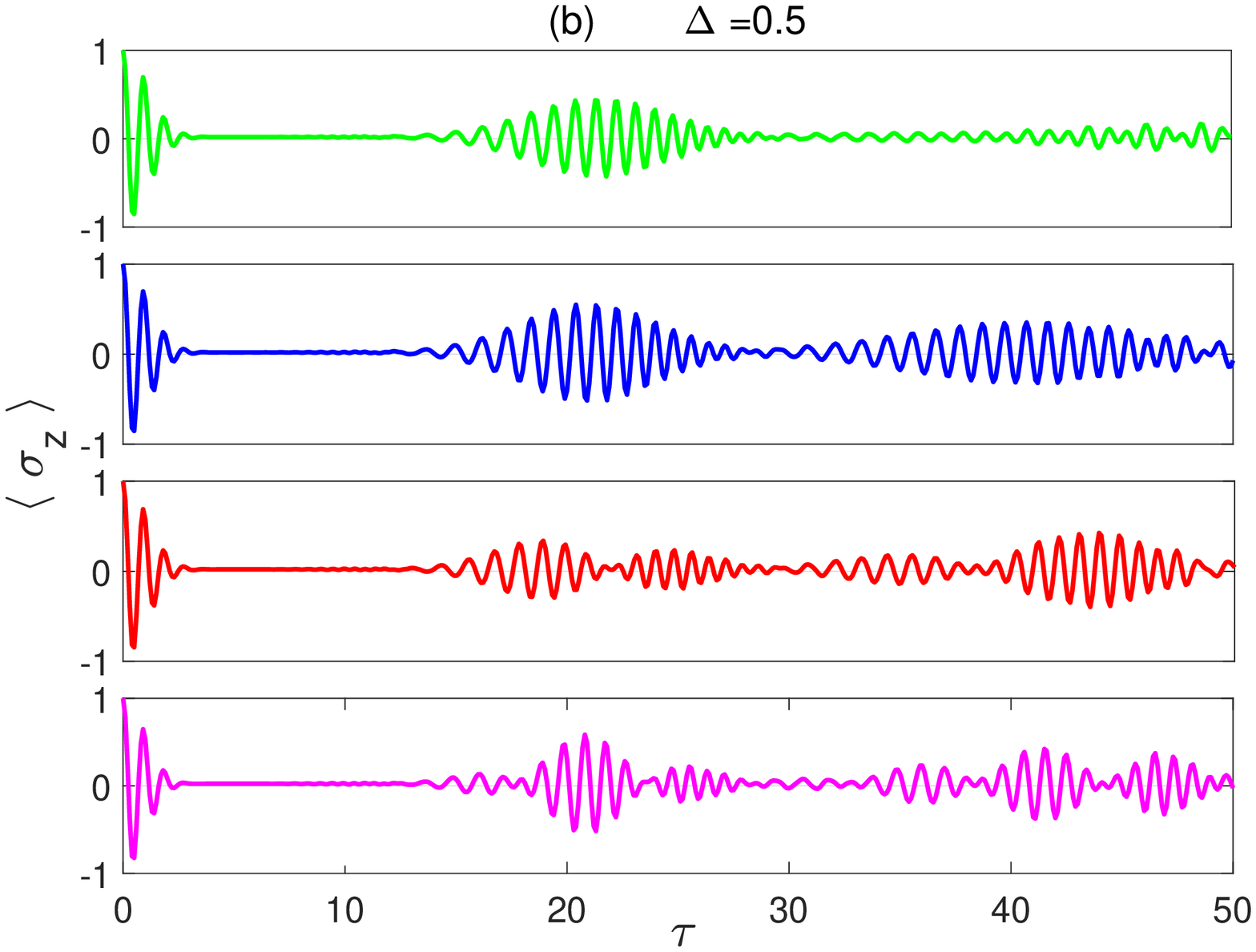}}\quad
\subfigure{\includegraphics[width=9cm]{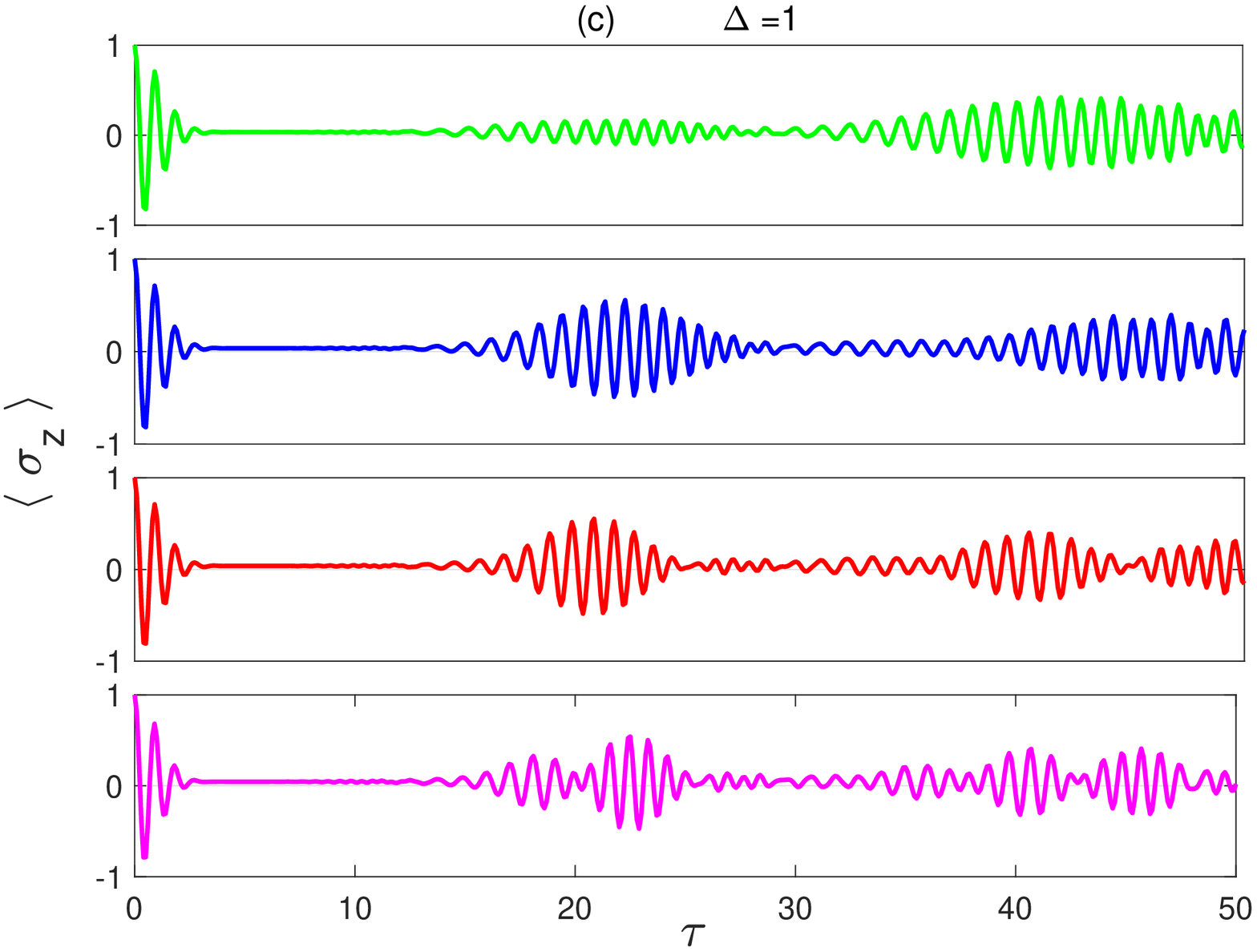}} \quad
\caption{{\protect\footnotesize (Color online) Dynamics of atomic inversion versus the scaled time $\tau=\lambda_1 t$ with the two quantum systems are initially  in a disentangled state $\vert\psi_{e}(0)\rangle=\vert e_{1}\rangle \vert e_{2}\rangle$ and the field is in a coherent state for various parameter values: $\lambda_{2}=0,0.1,0.5,1$ (in units of $\lambda_1$), detuning parameter $\Delta=0,0.5,1$ and mean number of photons $\bar{n}=10$. The legend is as shown in panel (a).}}
\label{up_n=10}
\end{minipage}
\end{figure}
\begin{figure}[htbp]
\begin{minipage}[c]{\textwidth}
\subfigure{\includegraphics[width=9cm]{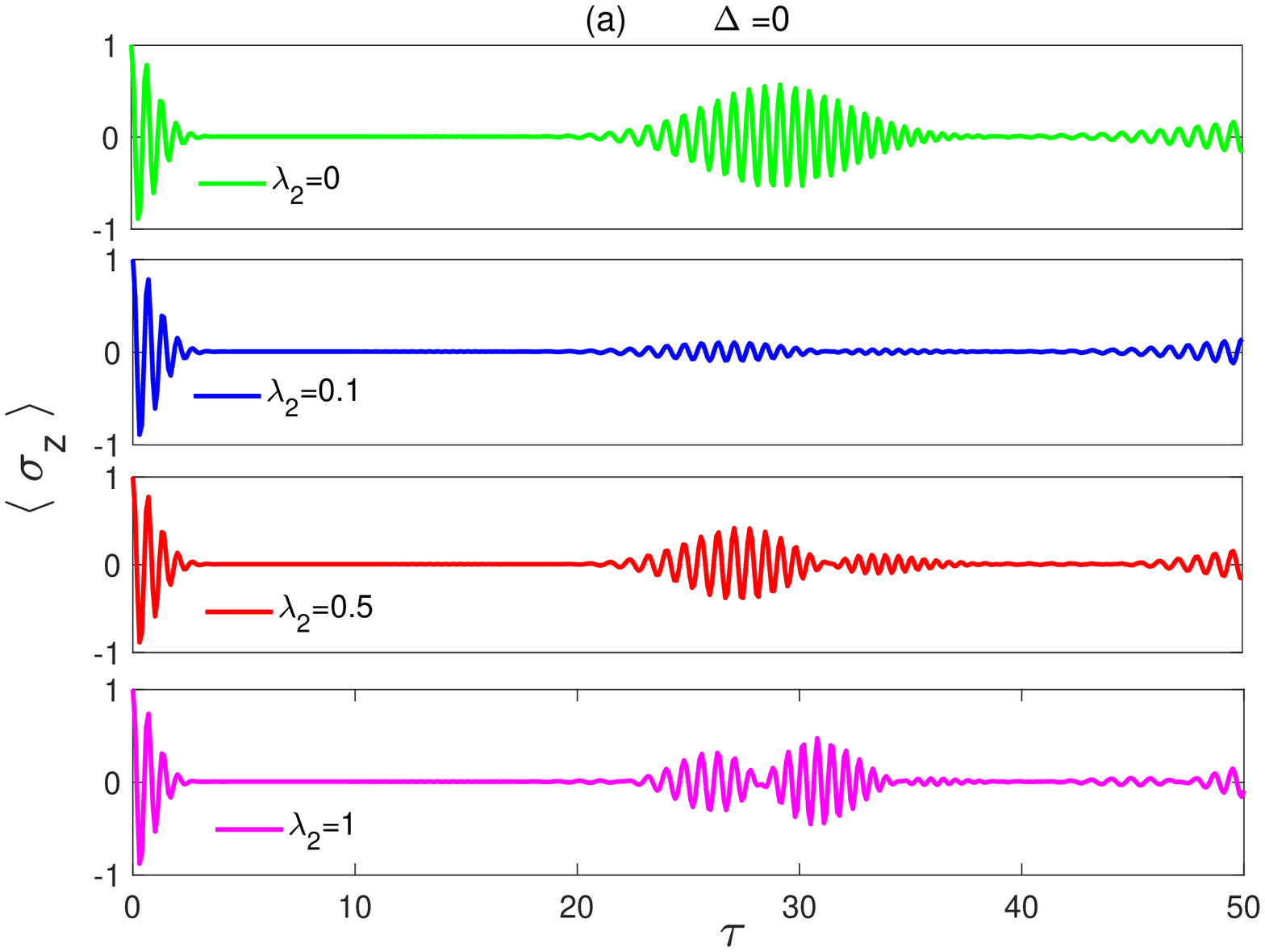}} \quad
\subfigure{\includegraphics[width=9cm]{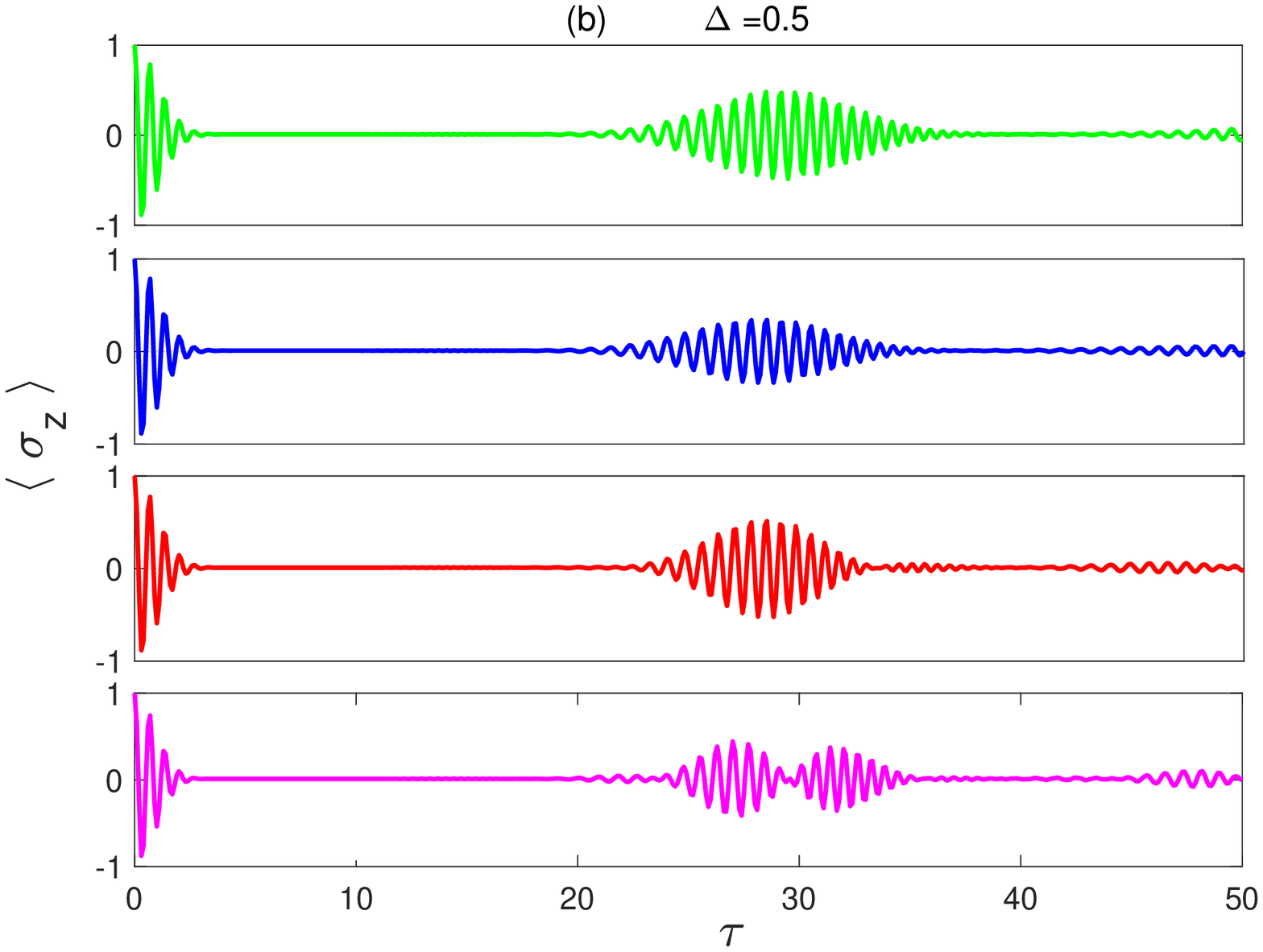}}\quad
\subfigure{\includegraphics[width=9cm]{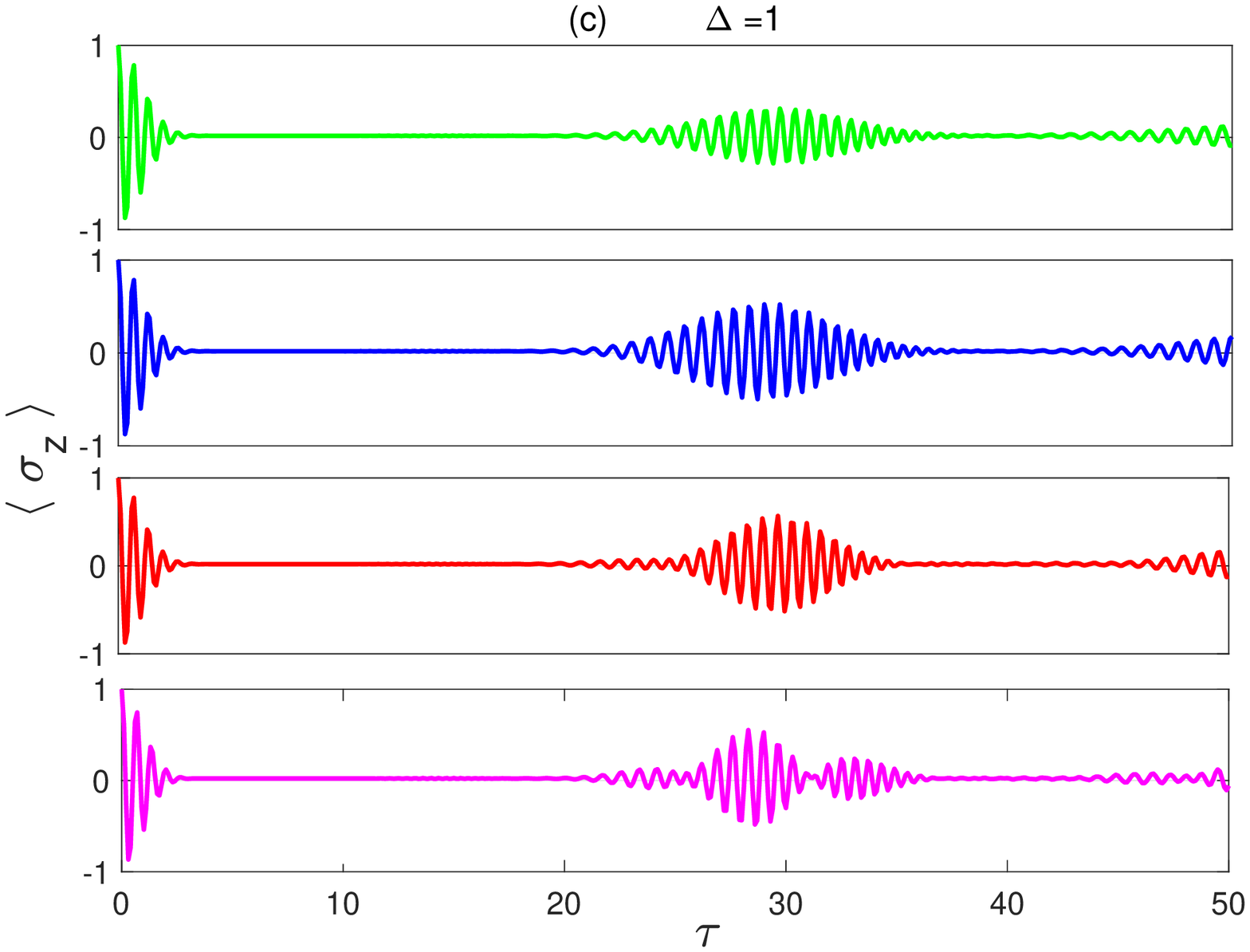}}\quad
\caption{{\protect\footnotesize (Color online) Dynamics of atomic inversion versus the scaled time $\tau=\lambda_1 t$ with the two quantum systems are initially  in a disentangled state $\vert\psi_{e}(0)\rangle=\vert e_{1}\rangle \vert e_{2}\rangle$ and the field is in a coherent state for various parameter values: $\lambda_{2}=0,0.1,0.5,1$ (in units of $\lambda_1$), detuning parameter $\Delta=0,0.5,1$ and mean number of photons $\bar{n}=20$. The legend is as shown in panel (a).}}
\label{up_n=20}
\end{minipage}
\end{figure}

\begin{figure}[htbp]
\begin{minipage}[c]{\textwidth}
\centering
\subfigure{\includegraphics[width=12.5cm]{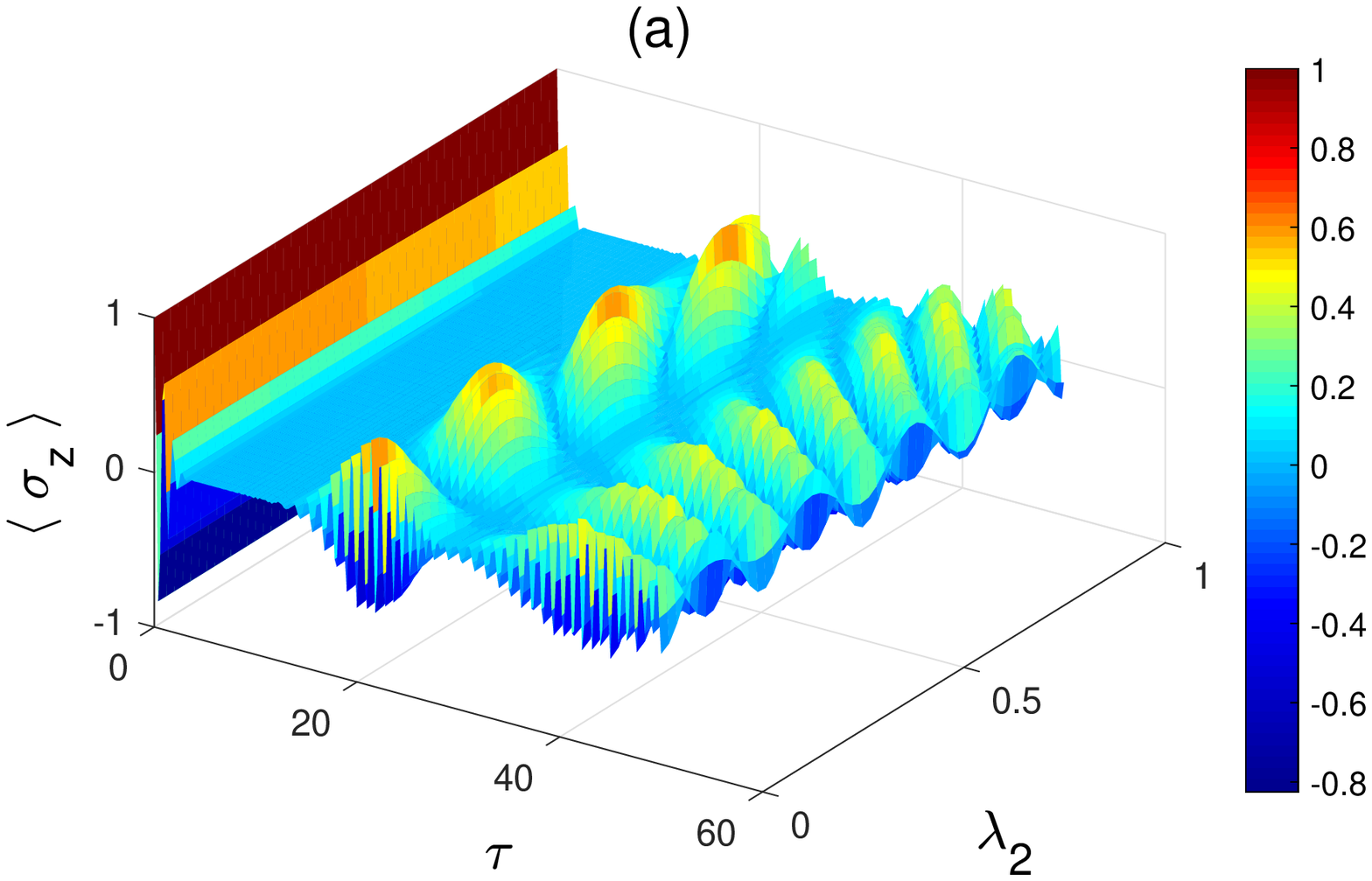}}\\
\subfigure{\includegraphics[width=12.5cm]{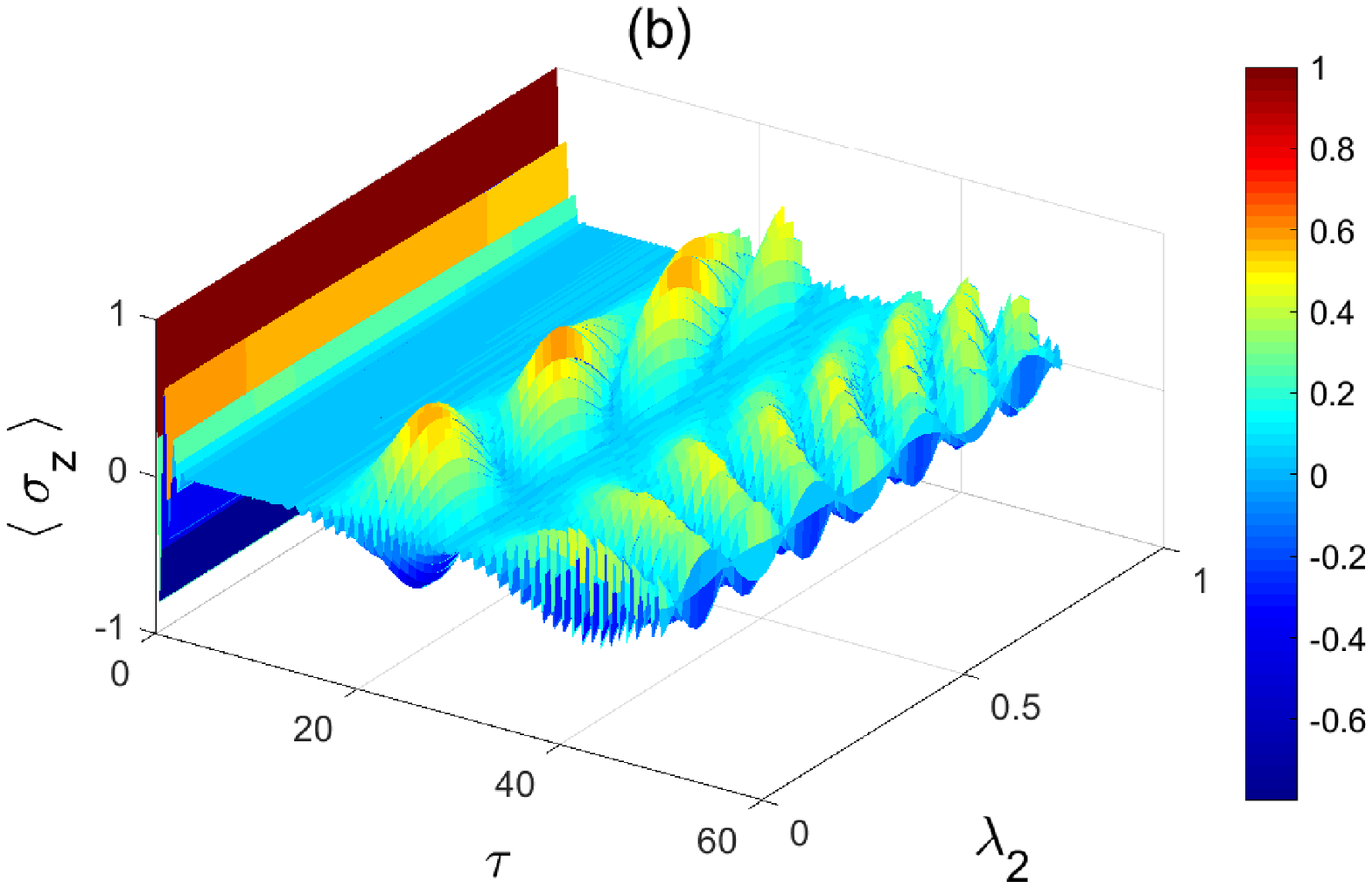}}
\caption{{\protect\footnotesize (Color online) Dynamics of atomic inversion versus the scaled time $\tau=\lambda_1 t$ and the coupling parameter $\lambda_2$ (in units of $\lambda_1$) with the two quantum systems are initially  in a disentangled state $\vert\psi_{e}(0)\rangle=\vert e_{1}\rangle \vert e_{2}\rangle$ and the field is in a coherent state with mean number of photons $\bar{n}=10$. The detuning parameter $\Delta=0$ in (a) and $1$ in (b).}}
\label{up_n=10_3D}
\end{minipage}
\end{figure}
\begin{figure}[htbp]
\begin{minipage}[c]{\textwidth}
\centering
\subfigure{\includegraphics[width=12.5cm]{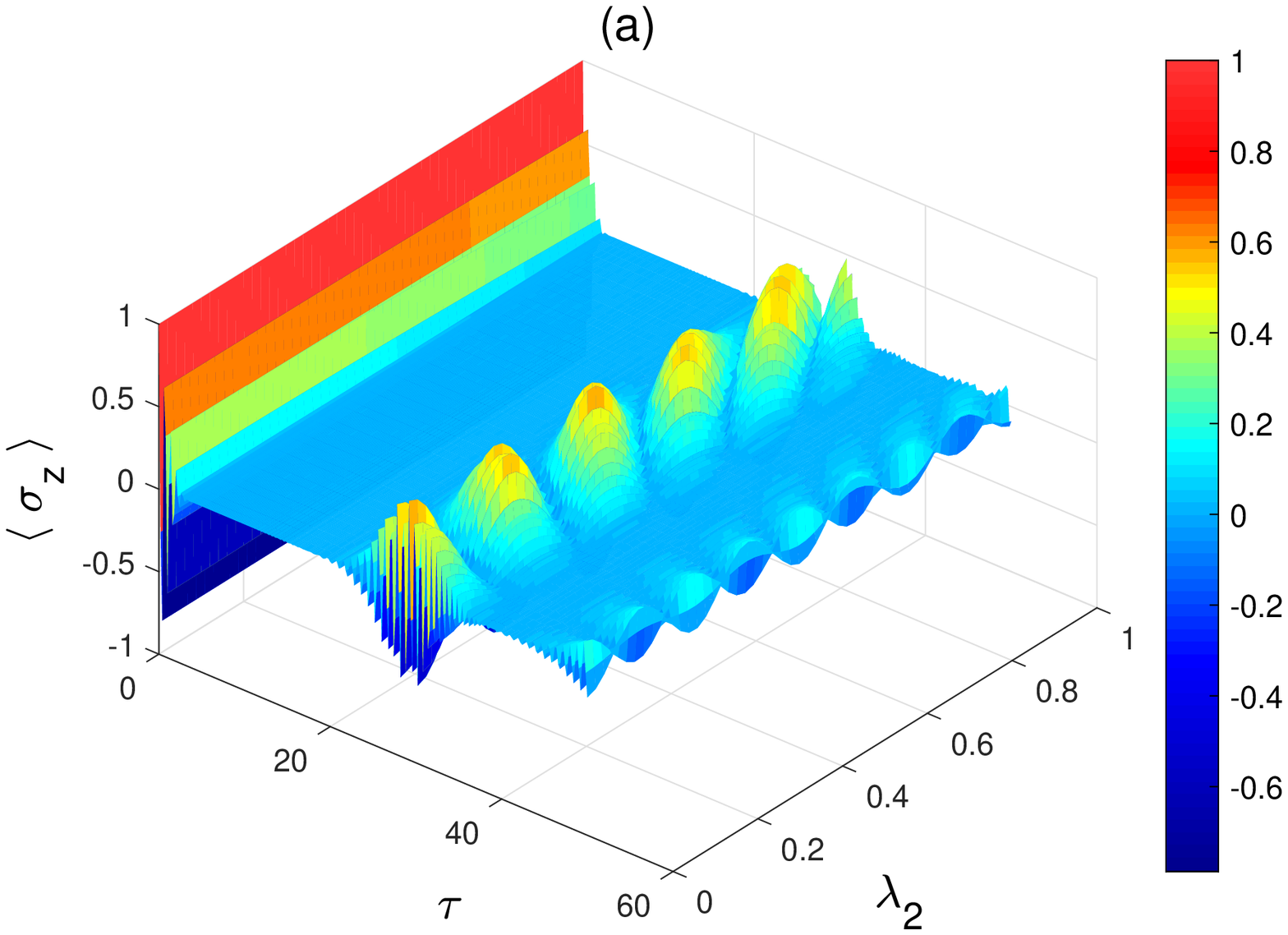}}\quad
\subfigure{\includegraphics[width=12.5cm]{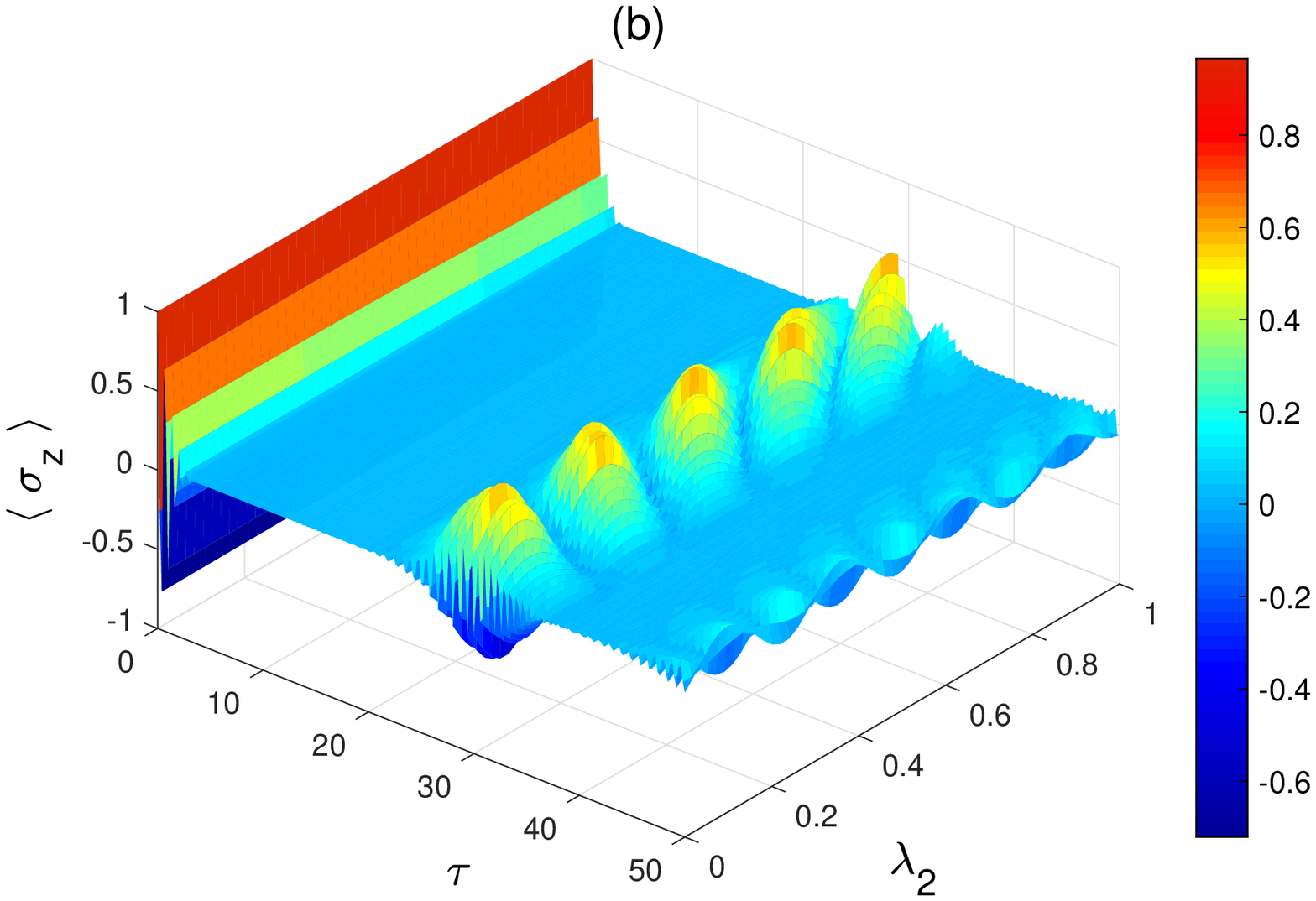}}\quad
\caption{{\protect\footnotesize (Color online) Dynamics of atomic inversion versus the scaled time $\tau=\lambda_1 t$ and the coupling parameter $\lambda_2$ (in units of $\lambda_1$) with the two quantum systems are initially  in a disentangled state $\vert\psi_{e}(0)\rangle=\vert e_{1}\rangle \vert e_{2}\rangle$ and the field is in a coherent state with mean number of photons $\bar{n}=20$. The detuning parameter $\Delta=0$ in (a) and $1$ in (b).}}
\label{up_n=20_3D}
\end{minipage}
\end{figure}
\begin{figure}[htbp]
\begin{minipage}[c]{\textwidth}
\centering
\subfigure{\includegraphics[width=12.5cm]{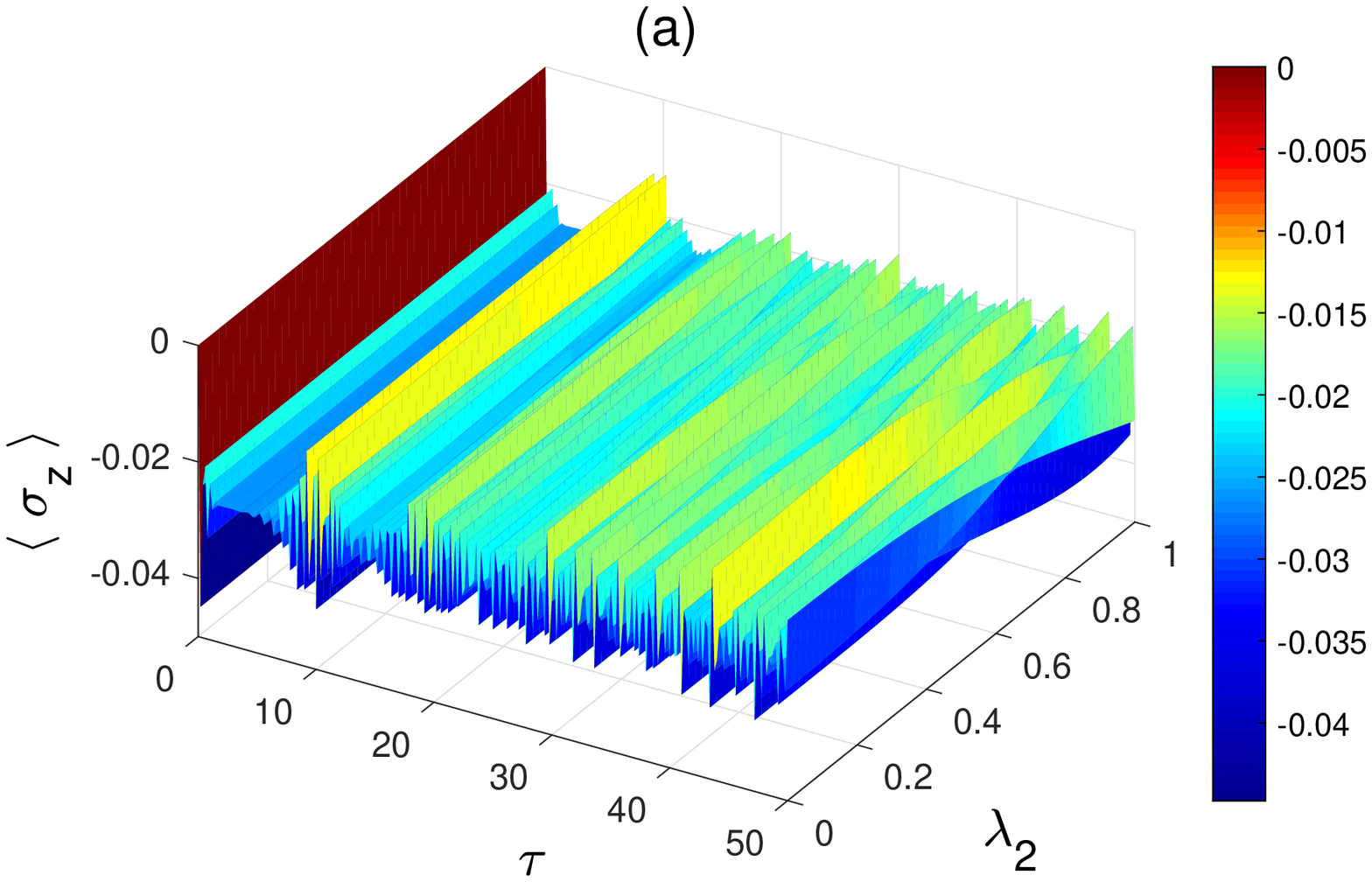}}\quad
\subfigure{\includegraphics[width=12.5cm]{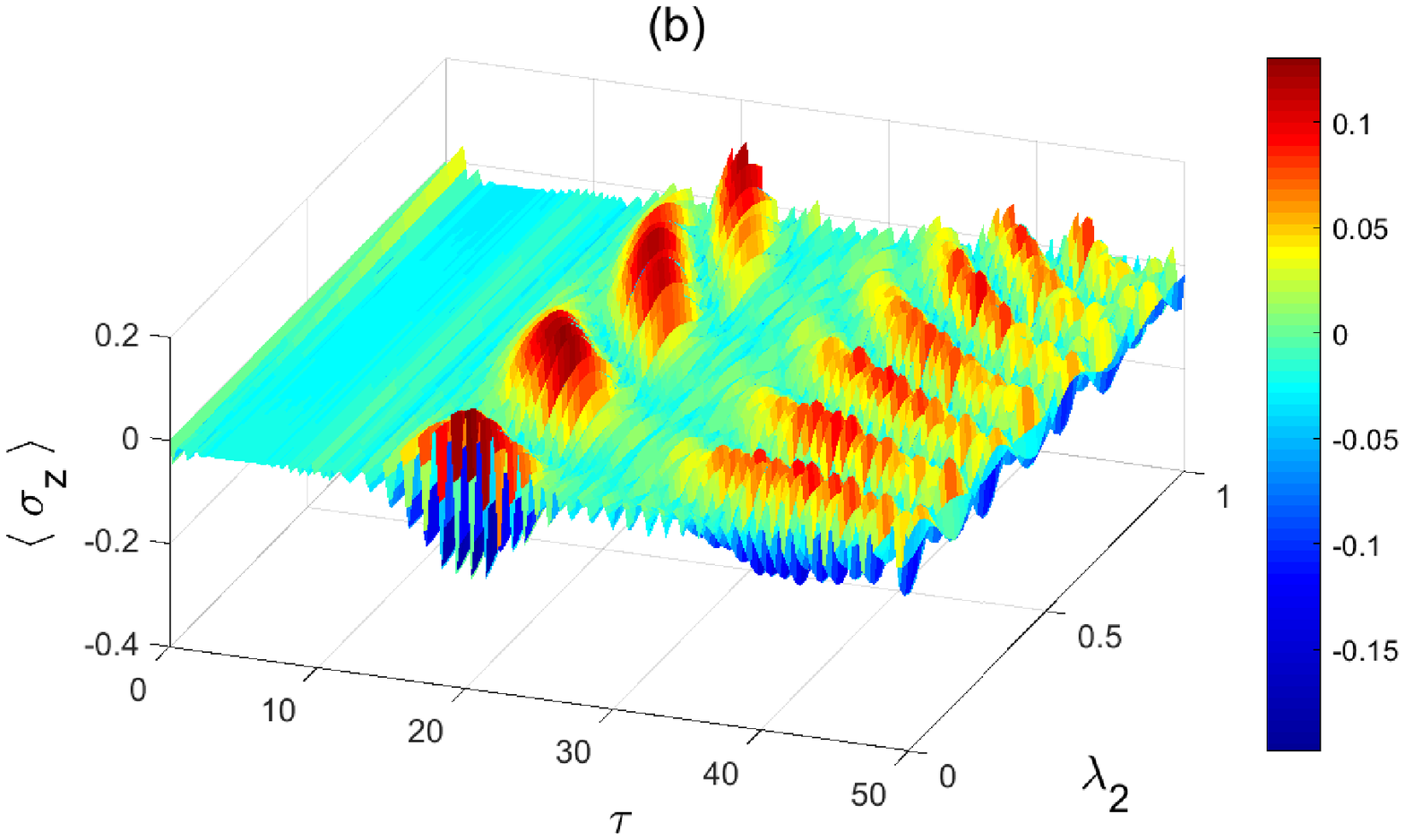}}\quad
\caption{{\protect\footnotesize (Color online) Dynamics of atomic inversion versus the scaled time $\tau=\lambda_1 t$ and the coupling parameter $\lambda_2$ (in units of $\lambda_1$) with the two quantum systems are initially in a maximally entangled Bell state $\vert\psi_{b}(0)\rangle=\frac{1}{\sqrt{2}}[\vert e_{1}\rangle \vert e_{2}\rangle + \vert g_{1}\rangle \vert g_{2}\rangle]$ and the field is in a coherent state with mean number of photons $\bar{n}=10$. The detuning parameter $\Delta=0$ in (a) and $1$ in (b).}}
\label{bell_n=10_3D}
\end{minipage}
\end{figure}

To investigate the atomic inversion we first calculate the reduced density matrix of any one of the two symmetric quantum systems, say the first, $\hat{\rho}_{s_1}(t)$ by tracing out the other one in the two quantum systems reduced density matrix $\hat{\rho}_{qs}$ (Eq.~\ref{qs_rdm}), which leads to 
\begin{equation}
\hat{\rho}_{s_1}(t)=Tr_{s_2}\;\hat{\rho}_{qs}(t) =\displaystyle  \left(\begin{array}{cc}
\rho_{11} & \rho_{12} \\
\rho_{21} & \rho_{22}
\end{array}\right),
\label{eq:33}
\end{equation} 
where
\begin{eqnarray}
\nonumber \rho_{11}(t)&=&\sum_{n=0}^{\infty} \vert A_{n}(t) \vert^2 +  \vert B_{n+1}(t) \vert^2, \\
\nonumber \rho_{22}(t)&=& \sum_{n=0}^{\infty}\vert C_{n+1}(t) \vert^2 + \vert D_{n+2}(t) \vert^2, \\
\rho_{12}(t)&=& \rho^{*}_{21}(t)= \sum_{n=0}^{\infty} A_{n+1}(t)C_{n+1}^{*}(t) + B_{n+2}(t)D_{n+2}^{*}(t).
\label{eq:rho_elements}
\end{eqnarray}
Therefore, for the first system 
\begin{eqnarray}
\nonumber\langle \hat{\sigma}_{z}(t) \rangle &=& Tr [\hat{\rho}_{s_1}(t)\hat{\sigma}_{z}]\\
&=& \sum_{n=0}^{\infty} \vert A_{n}(t) \vert^2 +  \vert B_{n+1}(t)\vert^2 - \vert C_{n+1}(t) \vert^2 - \vert D_{n+2}(t) \vert^2.
\label{eq:sigma_z} 
\end{eqnarray}
To discuss the collapse-revival phenomenon in the considered system, we plot $\langle \hat{\sigma}_{z}(t) \rangle$ against the scaled time $\tau =\lambda_{1}t$ for different values of the associated parameters in Figs.~\ref{up_n=10} to \ref{up_n=20_3D}. We assume that the two systems are initially in their excited upper levels, described by the disentangled state $\vert\psi_{e}(0)\rangle=\vert e_{1}\rangle \vert e_{2}\rangle$. The field is considered to be in a coherent state, which will be the case throughtout this paper. In Fig.~\ref{up_n=10}(a), we consider a zero detunning $(\Delta=0)$ while the mean number of photons $\bar{n}=10$ in all parts of Fig.~\ref{up_n=10}. In the top panel of Fig.~\ref{up_n=10}(a), we reproduce the collapse-revival behavior for the case of no coupling between the two quantum systems $(\lambda_2=0)$, presented before in ref.~\cite{Iqbal1988}. Then we introduce the two systems coupling effect (in units of $\lambda_1$) $\lambda_2=0.1,0.5$ and $1$ in the succeeding panels of (a). As can be noticed, when the coupling between the two quantum systems is turned on with a small strength $(\lambda_2=0.1)$, illustrated in the second panel of (a), the amplitude of the revival oscillation is reduced and a second collapse period is observed. This means the interaction between the two quantum systems suppresses the exchange of energy between each quantum system and the radiation field. As the coupling strength between the two systems is increased, shown in the last two panels of (a), the revival oscillations split into smaller ones. It is important to mention that the original collapse revival time in all cases has not been significantly affected by the coupling of the two quantum systems.
In Fig.~\ref{up_n=10}(b), we explore the effect of a non-zero detuning along with the coupling between the two systems. As can be noticed in the top panel of Fig.~\ref{up_n=10}(b), for zero coupling between the two systems after the first revival the second revival has a very small amplitude, but it is restored as $\lambda_2$ is introduced. Moreover, the strong coupling between the two systems causes the revival wave packets to split into smaller ones with no further collapse periods afterwards. A very similar behavior is observed in Fig.~\ref{up_n=10}(c) for a detuning parameter $\Delta=1$, again as the two systems coupling strength is increased the revival wave packets are split into smaller ones with small amplitude.

In Fig.~\ref{up_n=20}, we explore the effect of the field intensity in presence of the coupling between quantum systems on the dynamical behavior of the atomic inversion by increasing the mean number of photons, $\bar{n}=20$, where all the other system parameters are exactly the same as in Fig.~\ref{up_n=10}. The usual expected effect of a larger number of photons can be noticed, where the revival time becomes longer with the same value regardless of the values of the other system parameters tested in Fig.~\ref{up_n=20}. Comparing the top two panels in Fig.~\ref{up_n=20}(a) where $\Delta=0$, one can see that when the coupling between the two quantum systems is turned on with a small strength, $\lambda_2=0.1$, the main effect was the suppression of the revival oscillation, similar to what was observed in Fig.~\ref{up_n=10}(a). As the coupling strength increases the revival oscillation splits into smaller ones and spreads out over longer period of time as illustrated in the bottom two inner panels of Fig.~\ref{up_n=20}(a). In Fig.~\ref{up_n=20} (c) and (d) we test the effect of a non-zero detuning parameter, $\Delta = 0.5$ and 1 respectively, where as can be seen it reduces the suppression effect of the two system mutual coupling on the revival oscillation. Furthermore, the revival oscillation splits but this time at larger strength  
of the coupling parameter $\lambda_2$ in comparison with the resonance case.

For a better insight about the interplay between the mutual coupling between the two quantum systems, the number of photons and the detuning parameter $\Delta$ and their effect on the atomic inversion dynamics we depict the atomic inversion versus $\lambda_2$ (in units of $\lambda_1$) and the scaled time $\tau$ for different values of $\Delta$ in Figs.~\ref{up_n=10_3D} and \ref{up_n=20_3D} for $\bar{n} = 10$ and $20$ respectively. As can be noticed in Fig.~\ref{up_n=10_3D}(a), where $\Delta=0$, the revival oscillations at different values of $\lambda_2$ have gap regions (collapse regions) among themselves. These gap regions are normal to the $\lambda_2$ axis at small values of $\lambda_2$ but become more and more tilted as $\lambda_2$ increases. In fact, this explains many of the different characteristics observed in the dynamical behavior of the atomic inversion in Figs.~\ref{up_n=10} and \ref{up_n=20}. Obviously, by tuning the coupling strength between the two quantum systems $\lambda_2$ we can control the general profile of the collapse revival oscillation, the amplitude, the splitting of the revival envelope and slightly the primary collapse time and the subsequent ones. The effect of a non-zero detuning parameter, $\Delta = 1$, is illustrated in Fig.~\ref{up_n=10_3D}(b). The main effect is shifting the revival oscillations toward higher values of $\lambda_2$ and as a result a very weak oscillation is formed at $\lambda_2=0$ and the number of gap regions is reduced. In Fig.~\ref{up_n=20_3D} we explore the effect of higher number of photons on the dynamical behavior of the atomic inversion at $\Delta = 0$ and $\Delta = 1$ in (a) and (b) respectively. The overall behavior is very similar to what we have observed in Fig.~\ref{up_n=10_3D} except the usual longer primary and secondary collapse times caused by the extra number of photons.

The effect of a different initial state is considered in Fig.~\ref{bell_n=10_3D}, where the two quantum systems are prepared in a maximally entangled Bell state $\vert\psi_{b}(0)\rangle=\frac{1}{\sqrt{2}}[\vert e_{1}\rangle \vert e_{2}\rangle + \vert g_{1}\rangle \vert g_{2}\rangle]$. The dynamics of the atomic inversion at resonance, $\Delta=0$, as illustrated in Fig.~\ref{bell_n=10_3D}(a), is completely different from the previous cases where after a constant value period the revival oscillation becomes very sharp with a negative average value and very high frequency with no second constant value period. Interestingly, for a nonzero detuning parameter, $\Delta = 1$ shown in Fig.~\ref{bell_n=10_3D}(b), the profile becomes closer to the one observed in the previous initial state case with consecutive collapse revival pattern.
\section{Entanglement dynamics}
\label{Entanglement dynamics}
In this section we investigate the dynamics of bipartite entanglement between the two quantum systems in presence of the radiation field. All the information we may need can be extracted from the reduced density matrix of the two quantum systems $\hat{\rho}_{\textrm{qs}}(t)$ given by Eq.~(\ref{qs_rdm}). The entanglement between the two quantum system can be quantified with the help of the concurrence function $C(\rho_{\textrm{qs}})$ as proposed by Wootters~\cite{Wootters1998}, which is related to the entanglement of formation $E_{f}$ through the formula
\begin{equation}
E_{f}(\rho_{\textrm{qs}})= \mathcal{E}(C(\rho_{\textrm{qs}})),
\label{eq:48}
\end{equation}
where $\mathcal{E}$ is defined as
\begin{equation}
\mathcal{E}(C(\rho_{\textrm{qs}}))= h\left(\frac{1+ \sqrt{1-C^{2}(\rho_{\textrm{qs}})}}{2} \right),
\label{eq:49}
\end{equation}
here $h$ is 
the Shannon entropy function
\begin{equation}
h(x)=-x \log_{2} x - (1-x)\log_{2} (1-x),  
\label{eq:50}
\end{equation}
and the concurrence can by calculated from
\begin{equation}
C(\rho_{\textrm{qs}})= \max\;[0,\varepsilon_{1}-\varepsilon_{2}-\varepsilon_{3}-\varepsilon_{4}],
\label{eq:51}
\end{equation}
The $\varepsilon_{i}$ arranged in decreasing order are the square root of the four eigenvalues of the non-Hermitian matrix 
\begin{equation}
R\equiv \rho_{\textrm{qs}}\tilde{\rho}_{\textrm{qs}},
\label{eq:52}
\end{equation}
Where $\tilde{\rho}_{\textrm{qs}}$ is the spin flipped state defined as
\begin{equation}
\tilde{\rho}_{\textrm{qs}}=(\hat{\sigma}_{y}\otimes\hat{\sigma}_{y})\rho^{*}_{\textrm{qs}}(\hat{\sigma}_{y}\otimes\hat{\sigma}_{y}),
\label{eq:53}
\end{equation} 
Here $\rho^{*}_{\textrm{qs}}$ is the complex conjugate of $\rho_{\textrm{qs}}$ and $\hat{\sigma}_{y}$ is the Pauli spin matrix in the $y$ direction.\\
Both of  $C(\rho_{\textrm{qs}})$ and $E_{f}(\rho_{\textrm{qs}})$ go from $0$ for a separable state to $1$ for a maximally entangled state.\\
\begin{figure}[htbp]
\begin{minipage}[c]{\textwidth}
 \centering 
\subfigure{\includegraphics[width=8cm]{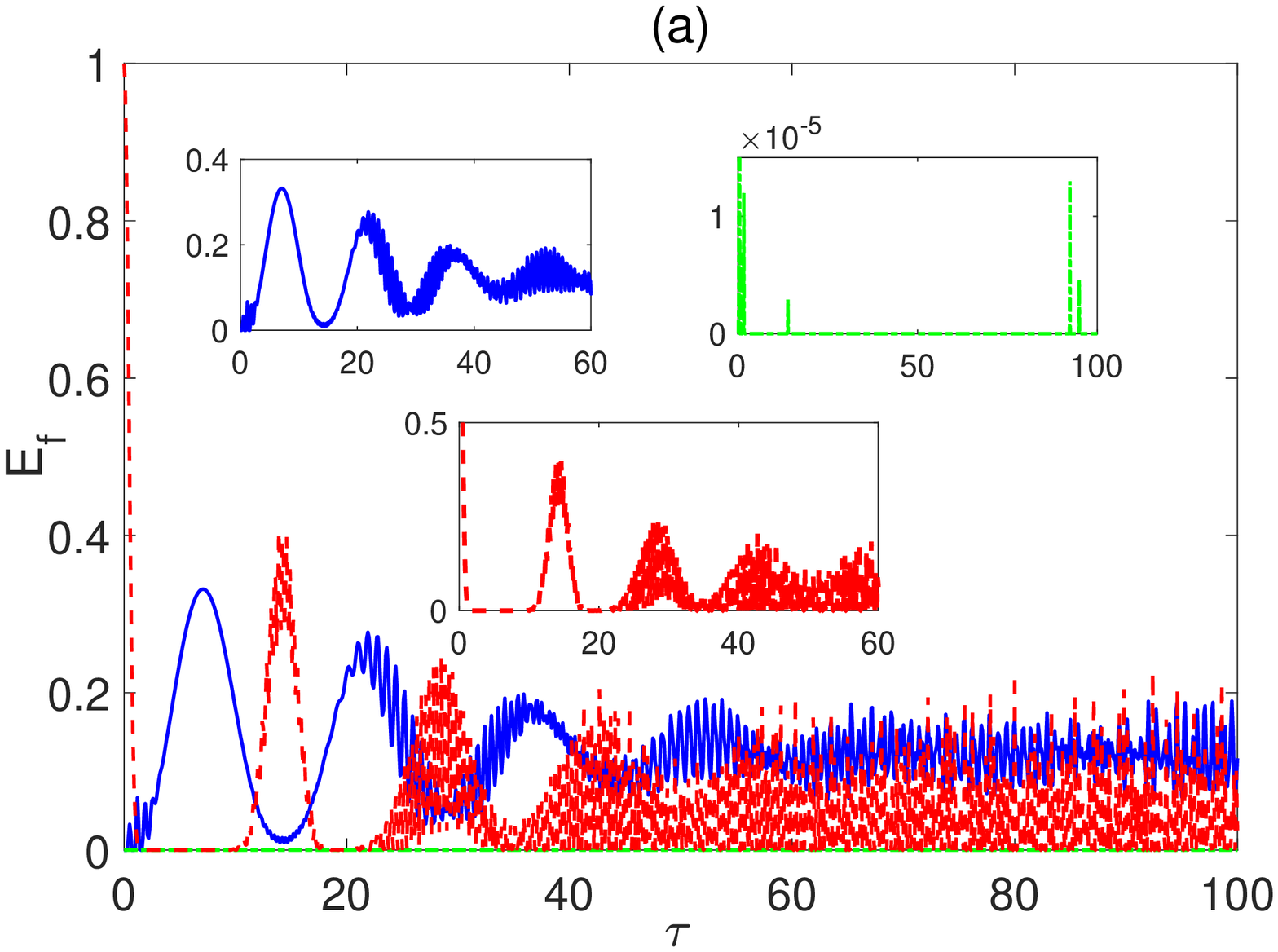}}\quad
   \subfigure{\includegraphics[width=8cm]{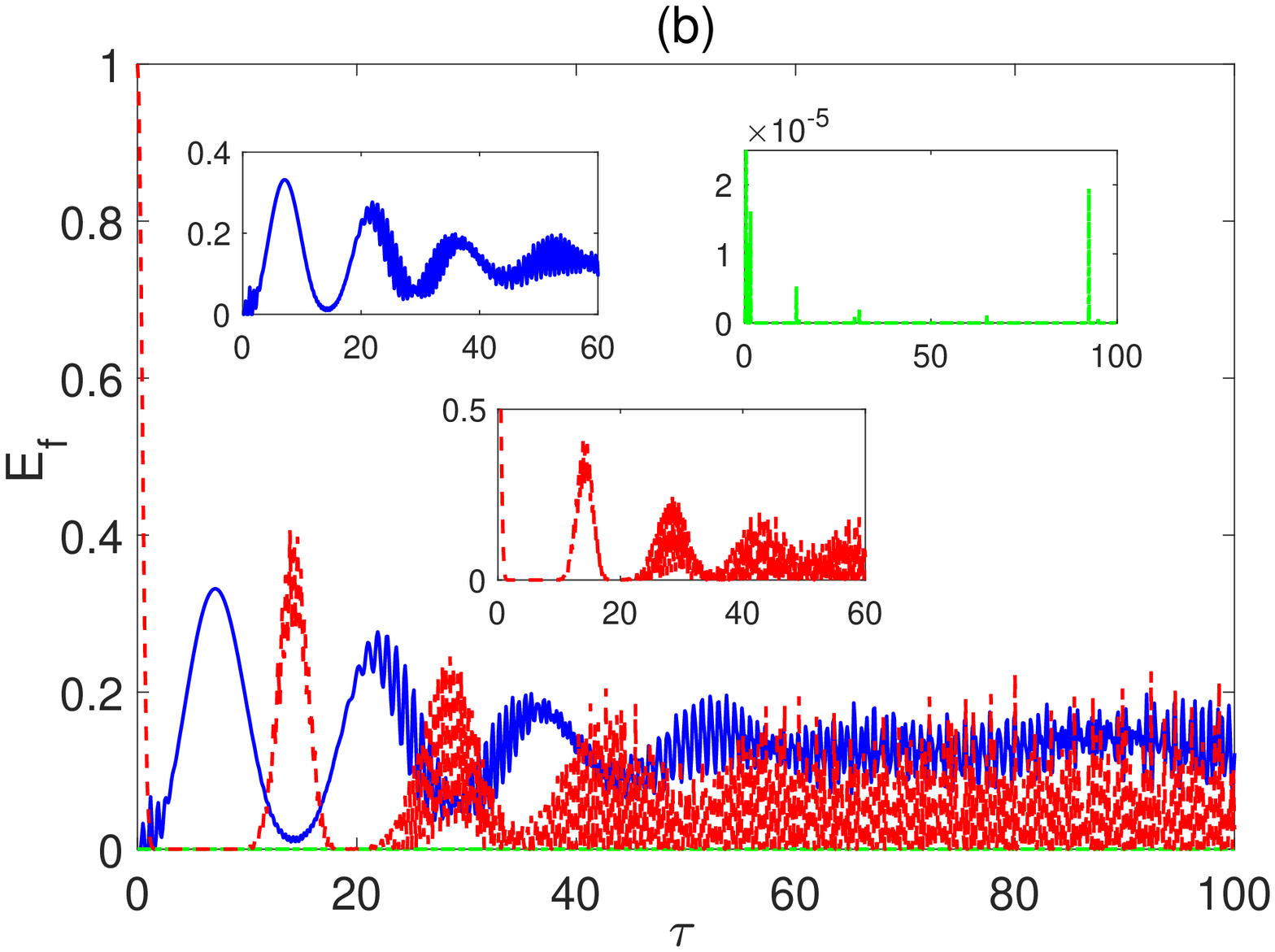}}\\
   \subfigure{\includegraphics[width=8cm]{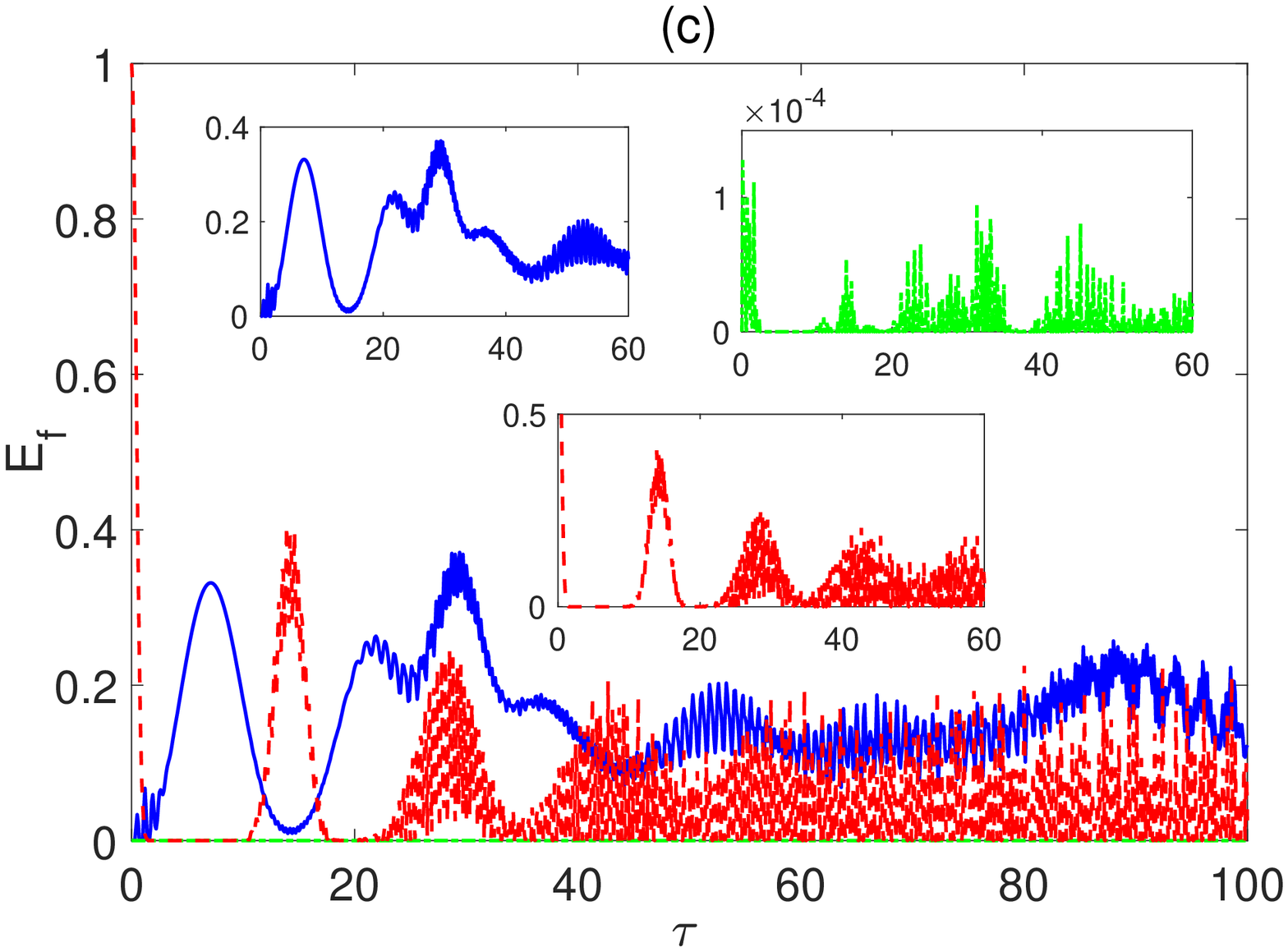}}\quad
   \subfigure{\includegraphics[width=8cm]{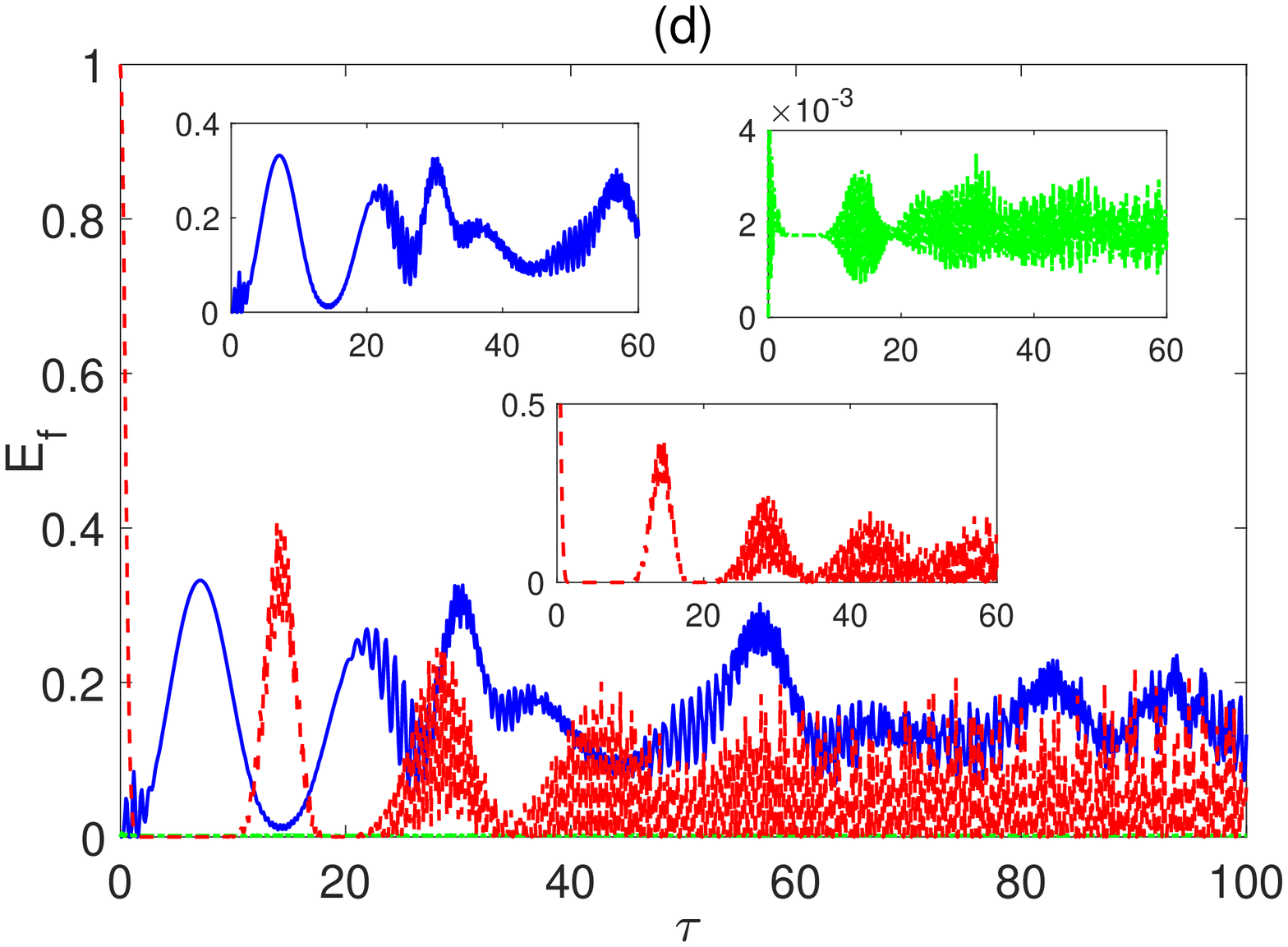}}\\   
    \caption{{\protect\footnotesize (Color online) Time evolution of entanglement versus the scaled time $\tau=\lambda_{1}t$ starting form different initial states at different values of the parameter $\lambda_{2}$ (in units of $\lambda_1$) (a) 0; (b) 0.01 ; (c) 0.1 and (d) 0.5, at zero detuning and $\bar{n}=20$ in all panels. The legend is as shown in panel 8(d).}}
\label{Ent_Lam1_1_Del_0}
 \end{minipage}
\end{figure}
\begin{figure}[htbp]
\begin{minipage}[c]{\textwidth}
 \centering
   \subfigure{\includegraphics[width=8cm]{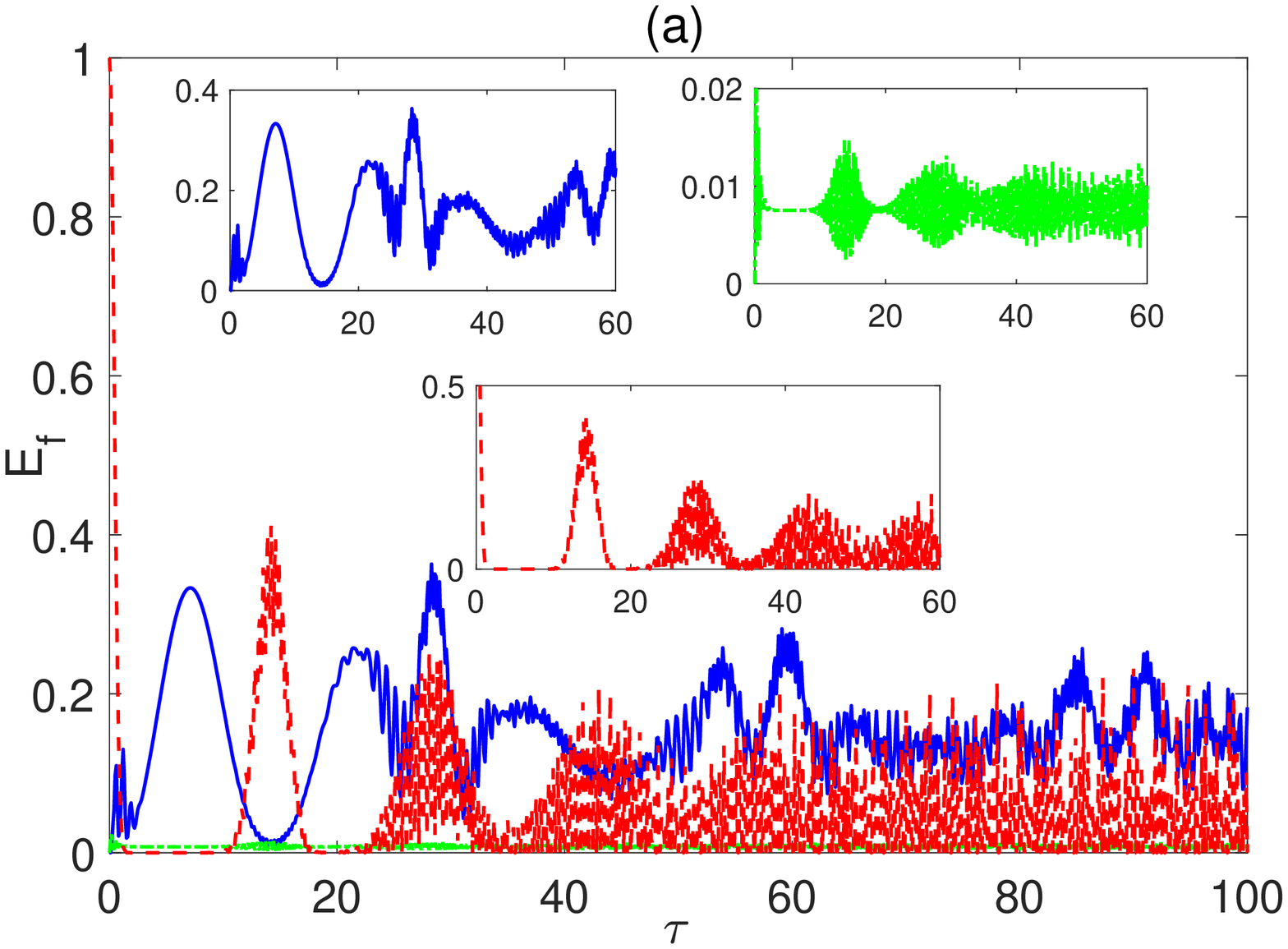}}\quad
   \subfigure{\includegraphics[width=8cm]{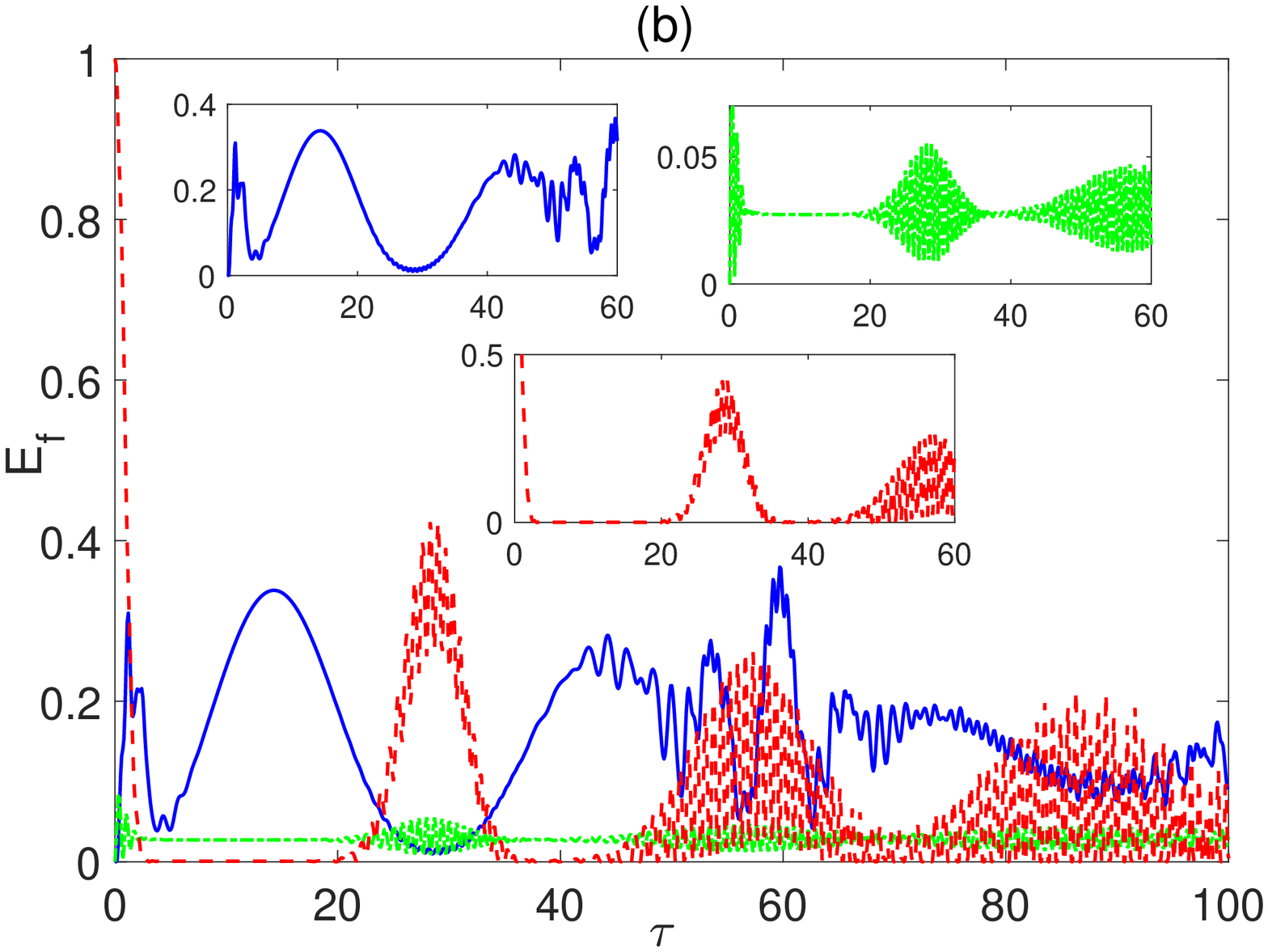}}\\
   \subfigure{\includegraphics[width=8cm]{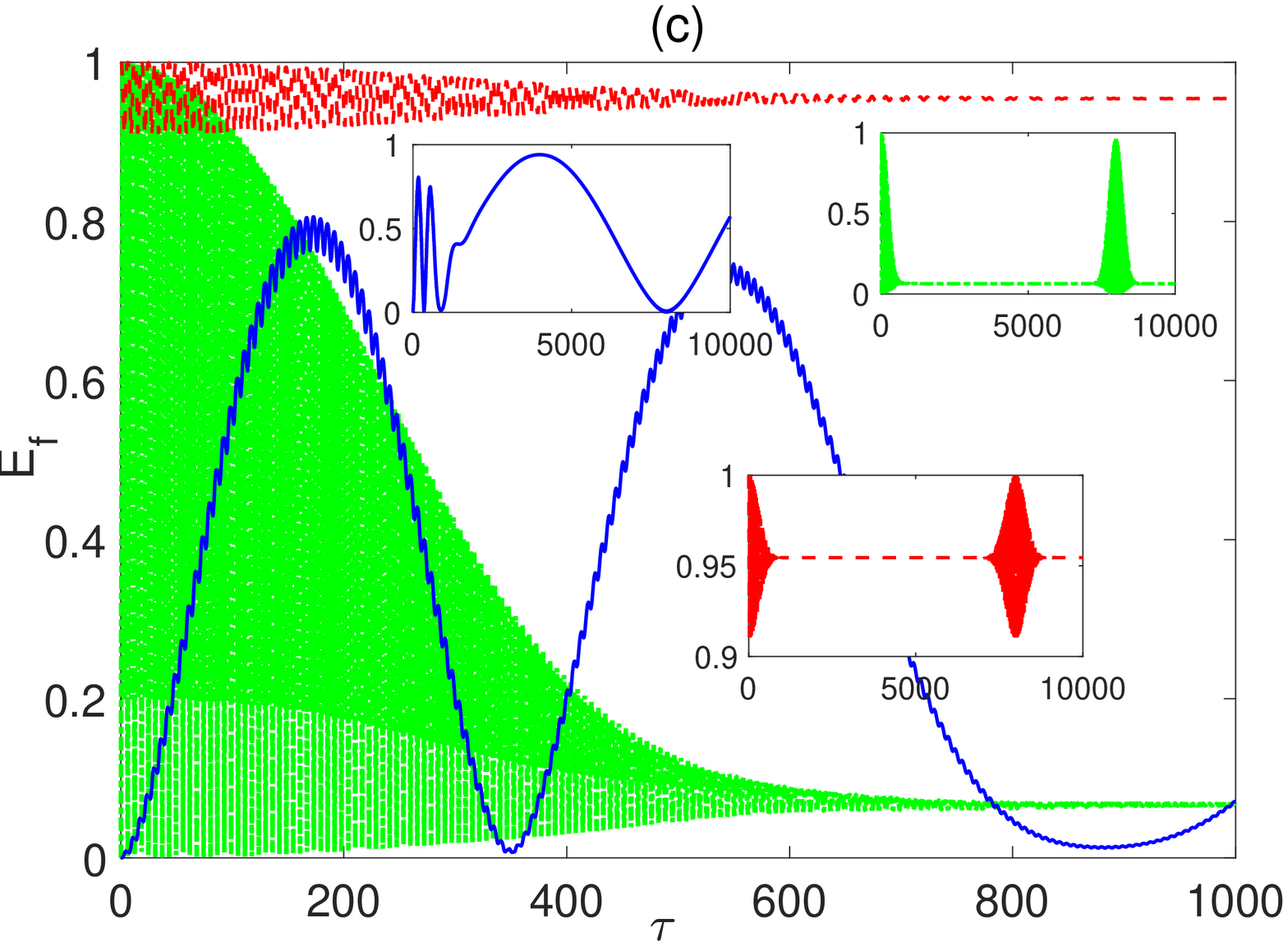}}\quad  
   \subfigure{\includegraphics[width=8cm]{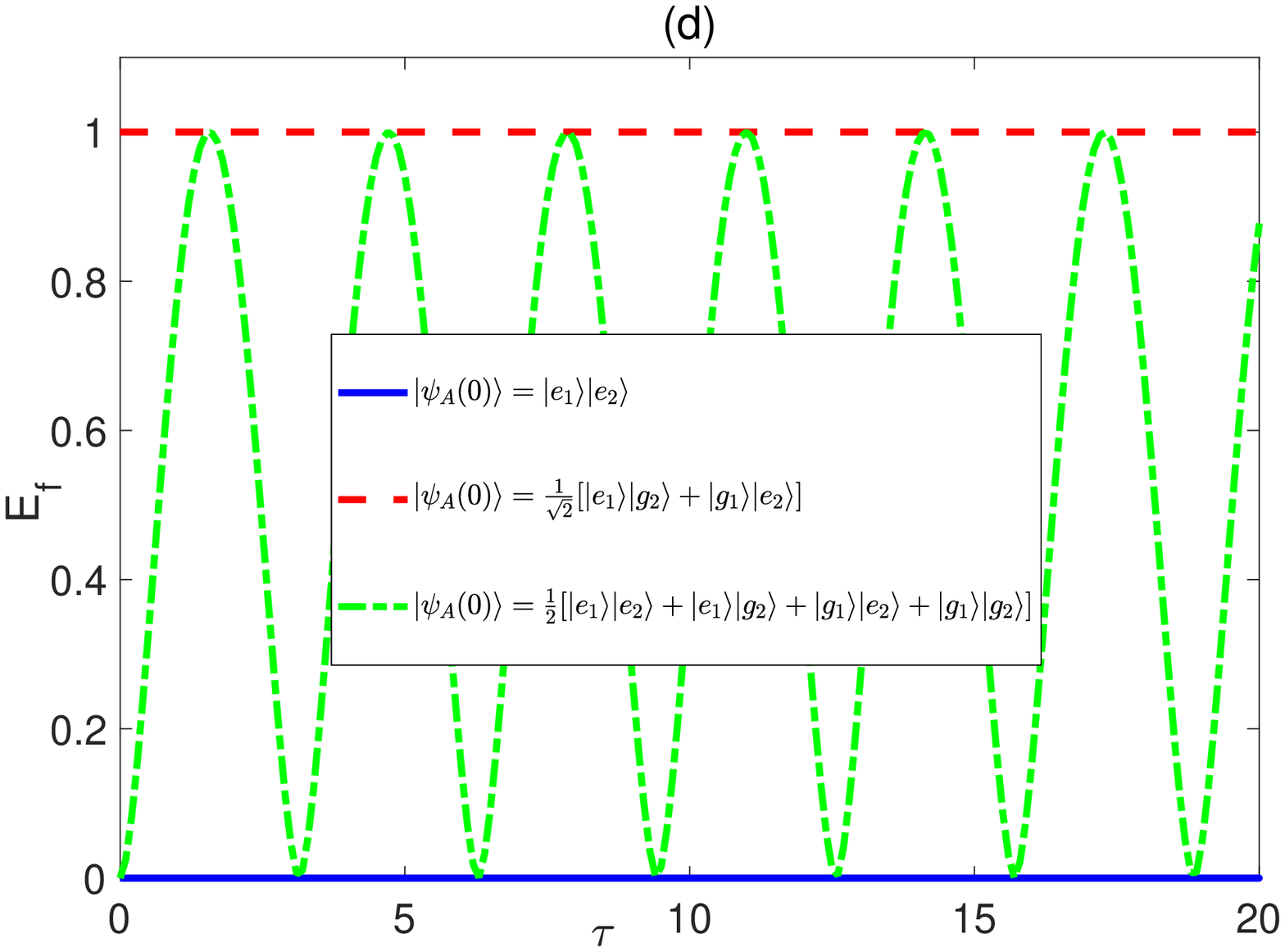}}\\   
  \caption{{\protect\footnotesize (Color online) Time evolution of entanglement versus the scaled time $\tau=\lambda_{2}t$ starting form different initial states at different values of the parameter $\lambda_{1}$ (in units of $\lambda_2$) (a) 1; (b) 0.5 ; (c) 0.01 and (d) 0, at zero detuning and $\bar{n}=20$ in all panels. The legend is as shown in panel (d).}}
\label{Ent_Lam2_1_Del_0}
 \end{minipage}
\end{figure}
\begin{figure}[htbp]
\begin{minipage}[c]{\textwidth}
 \centering 
\subfigure{\includegraphics[width=8cm]{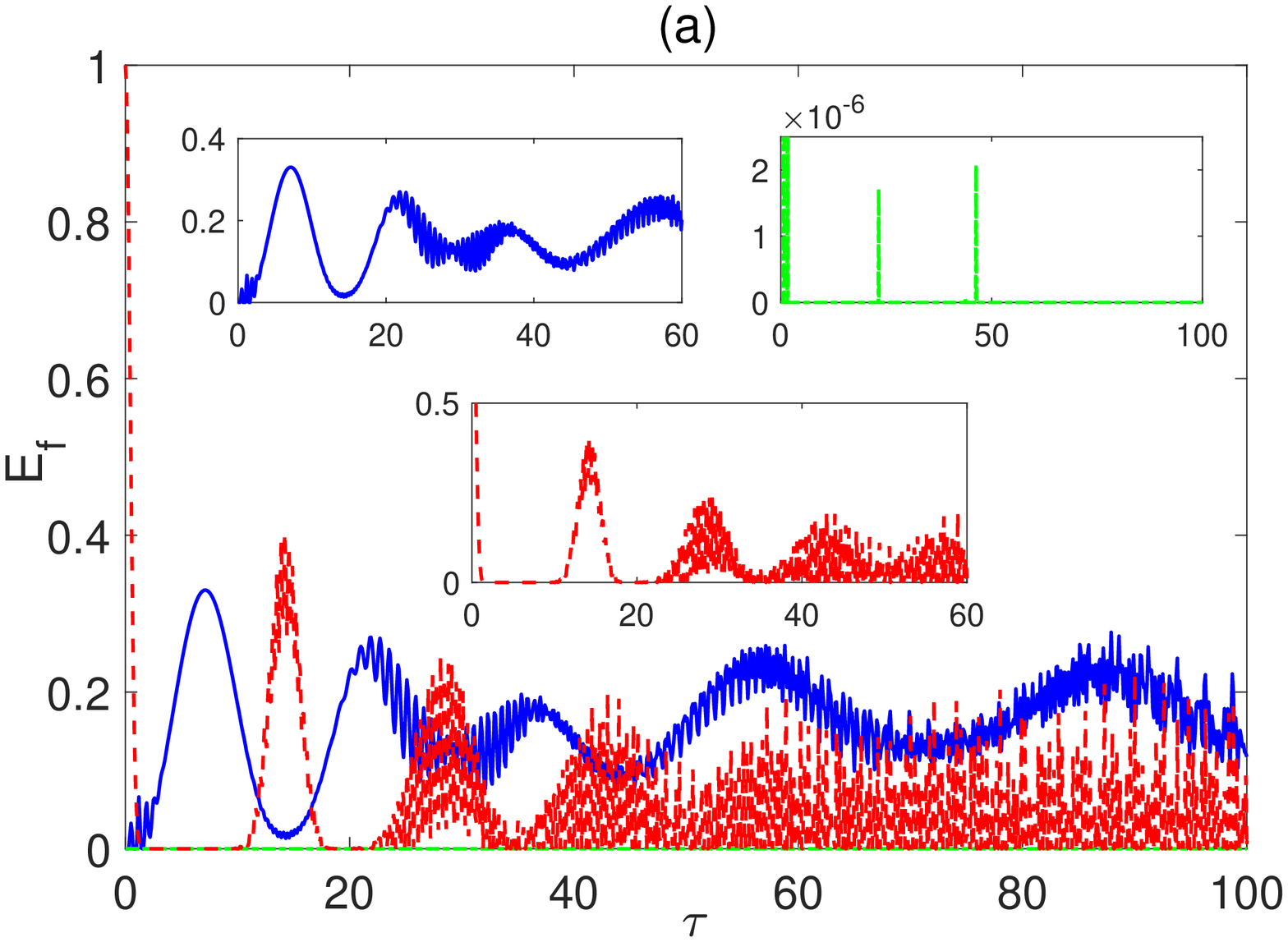}}\quad
   \subfigure{\includegraphics[width=8cm]{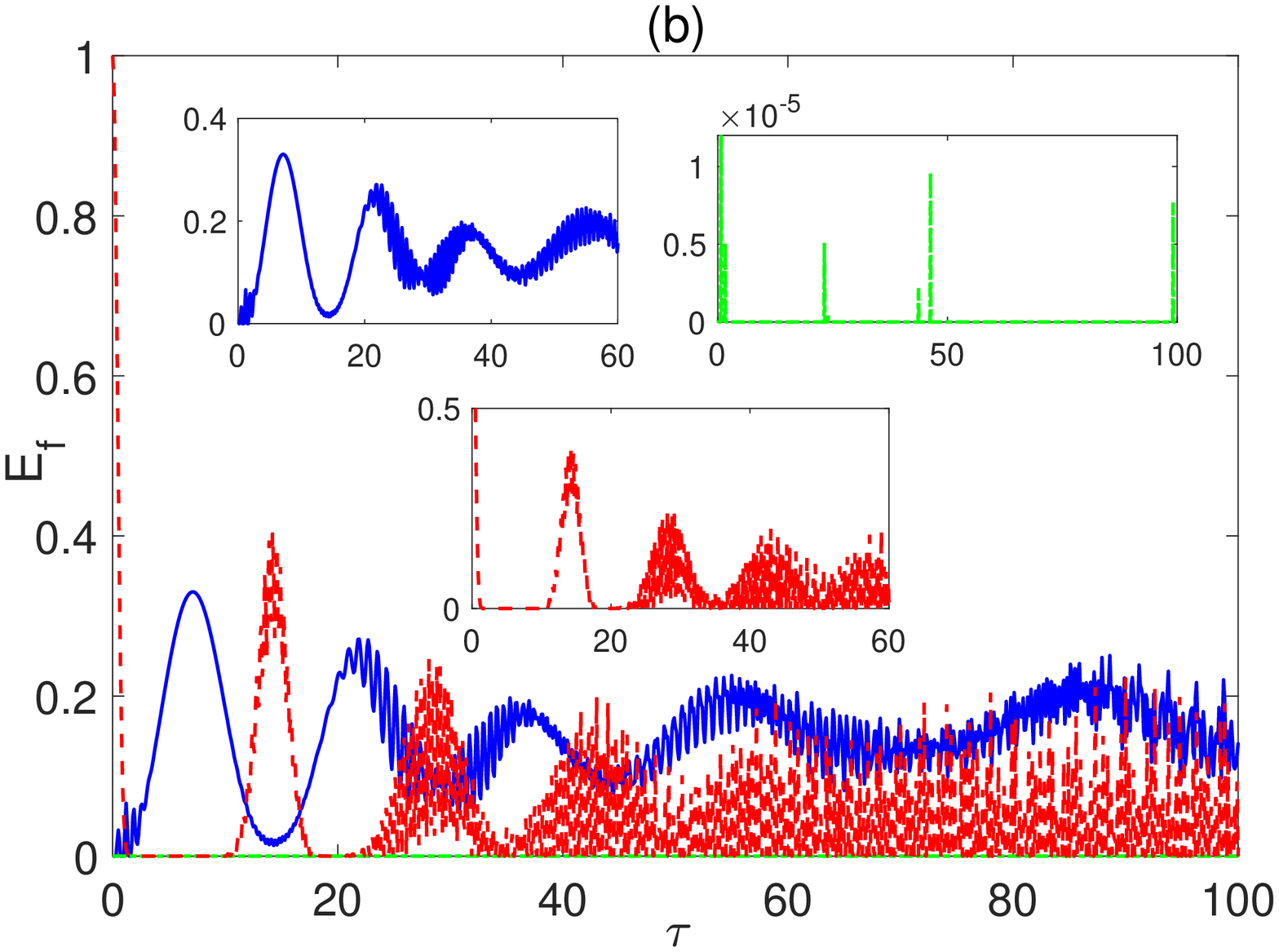}}\\
   \subfigure{\includegraphics[width=8cm]{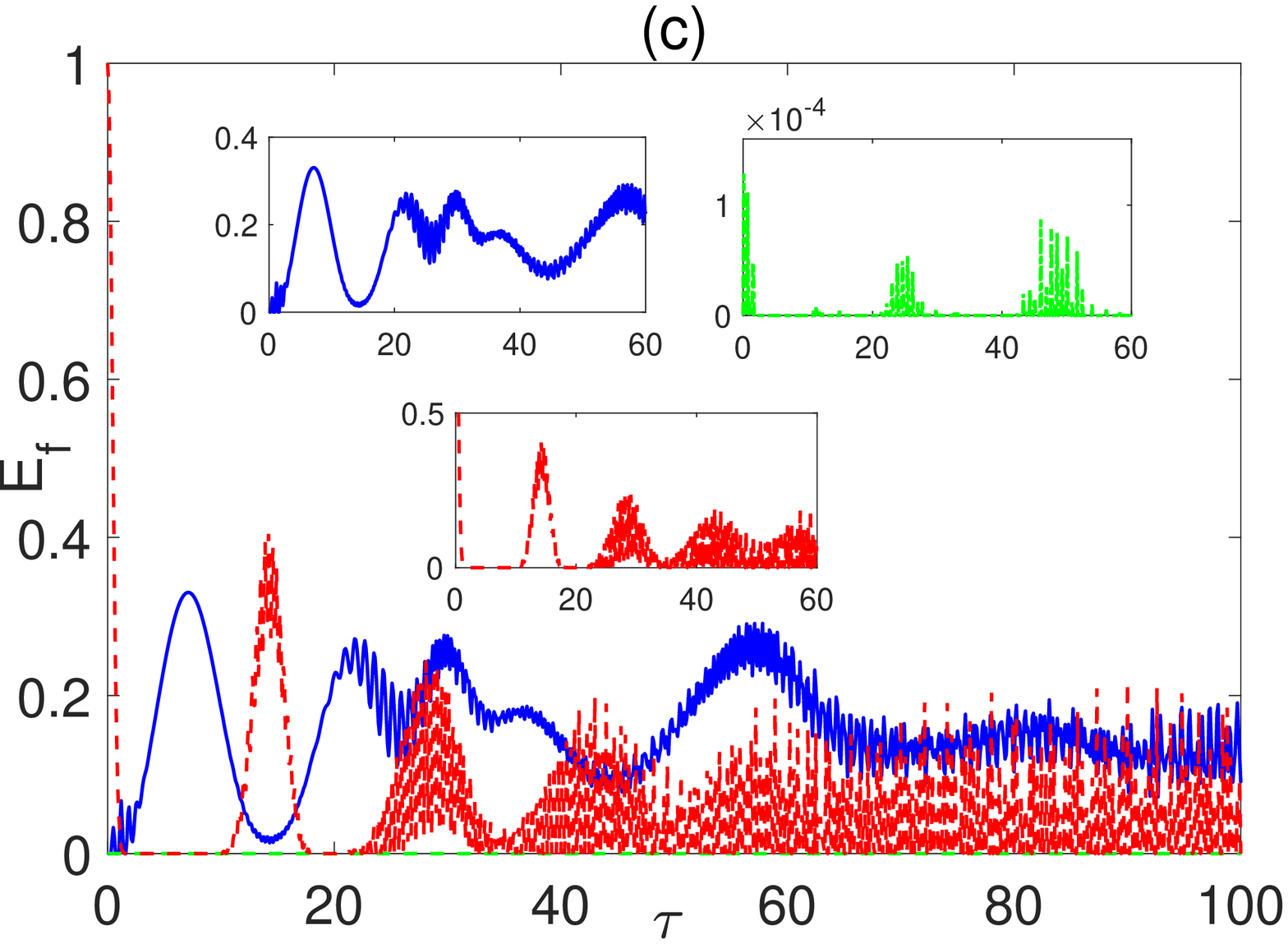}}\quad
   \subfigure{\includegraphics[width=8cm]{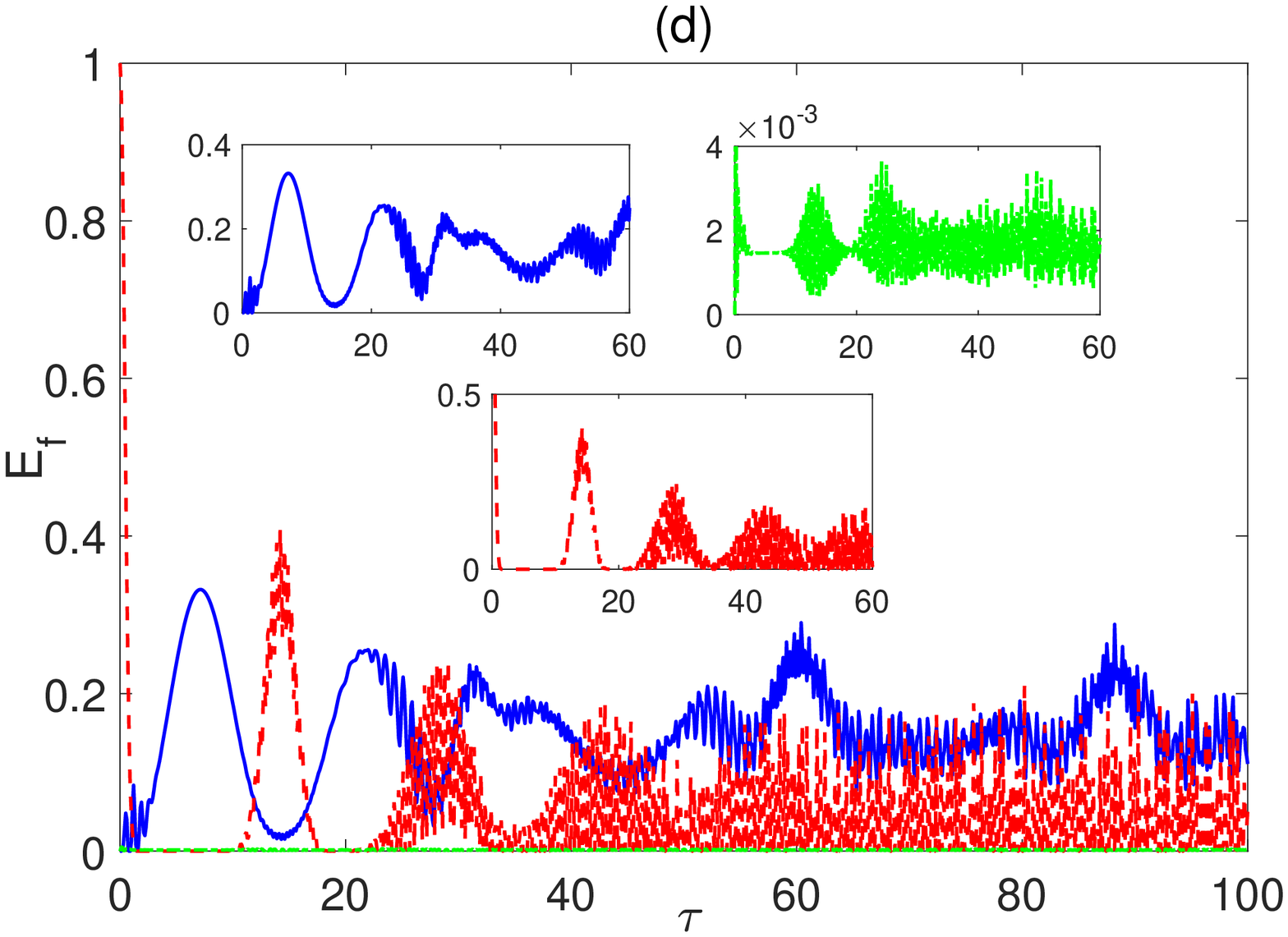}}\\   
    \caption{{\protect\footnotesize (Color online) Time evolution of entanglement versus the scaled time $\tau=\lambda_{1}t$ starting form different initial states at different values of the parameter $\lambda_{2}$ (in units of $\lambda_1$) (a) 0; (b) 0.01 ; (c) 0.1 and (d) 0.5, with detuning $\Delta=0.5$ and $\bar{n}=20$ in all panels. The legend is as shown in panel 8(d).}}
\label{Ent_Lam1_1_Del_05}
 \end{minipage}
\end{figure}
\begin{figure}[htbp]
\begin{minipage}[c]{\textwidth}
 \centering 
\subfigure{\includegraphics[width=8cm]{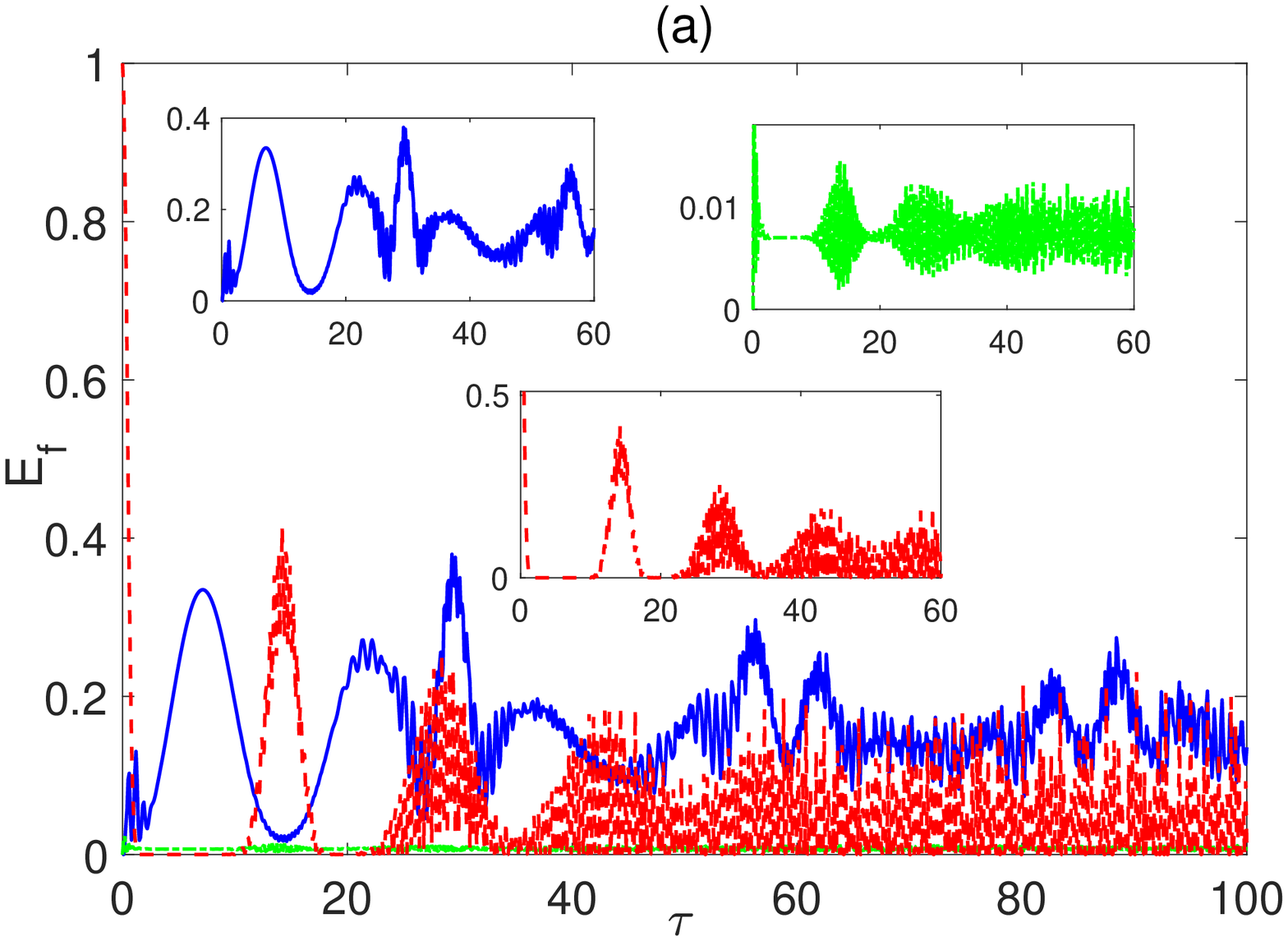}}\quad
   \subfigure{\includegraphics[width=8cm]{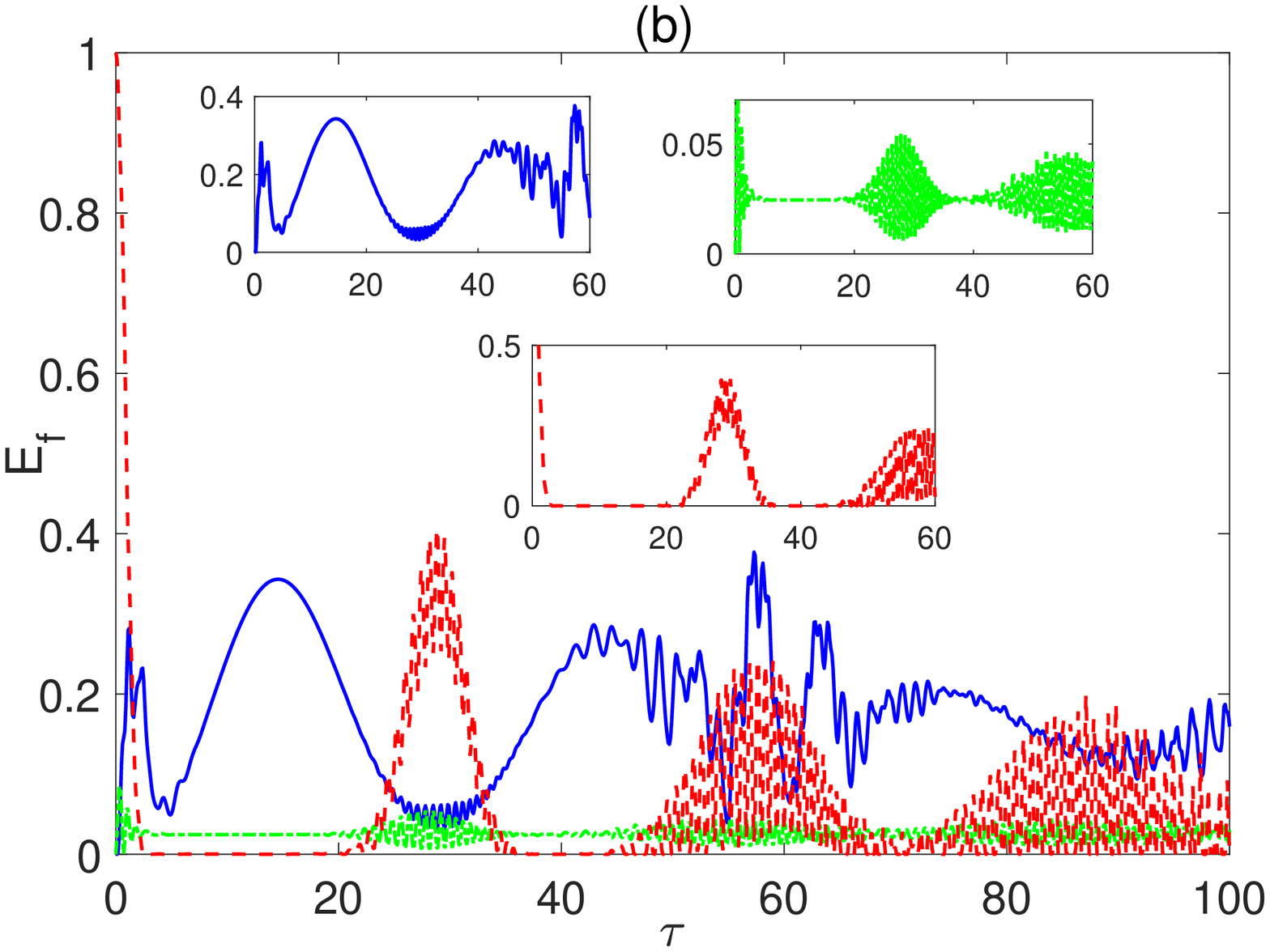}}\\
   \subfigure{\includegraphics[width=8cm]{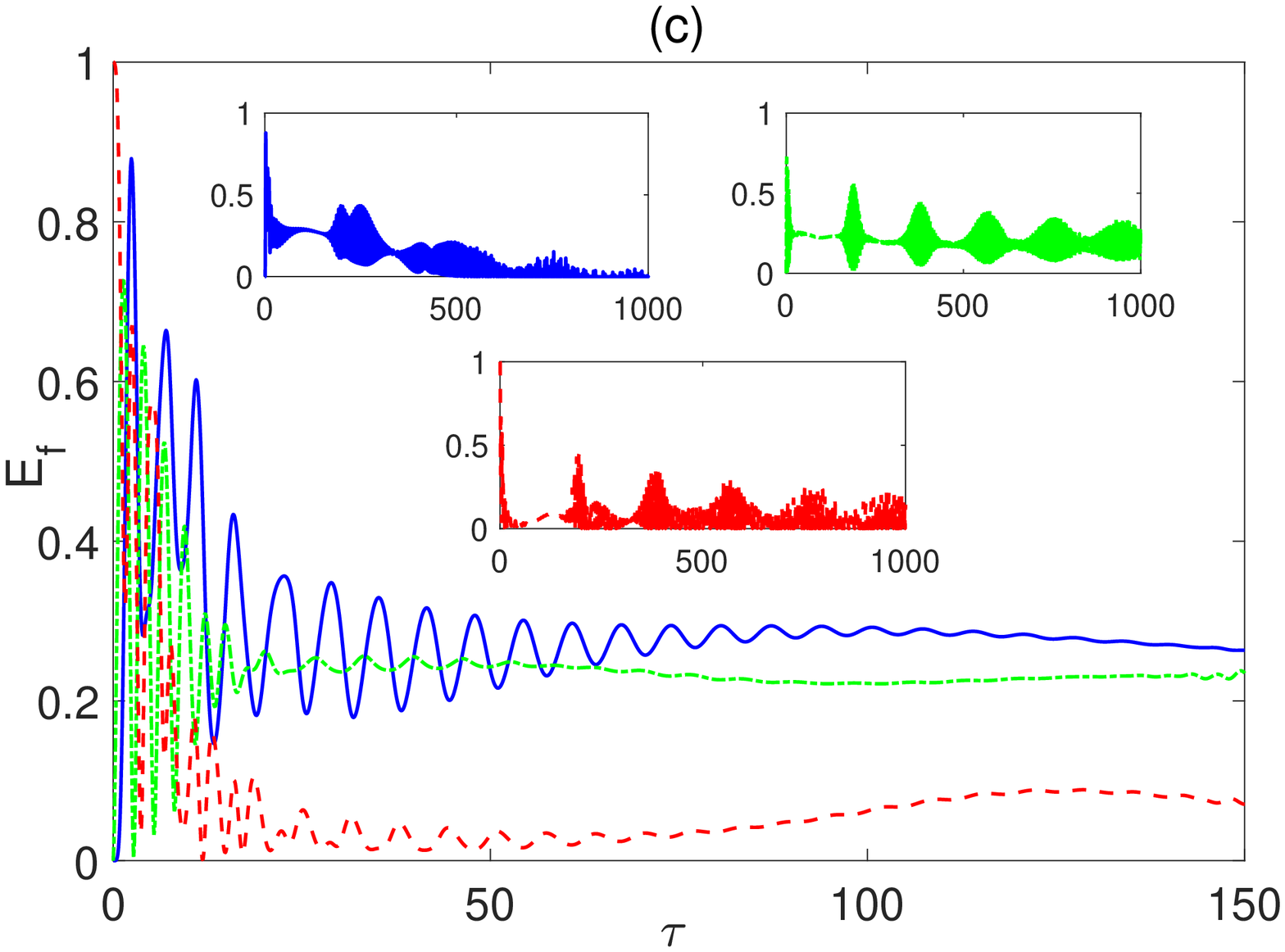}}\quad  
   \subfigure{\includegraphics[width=8cm]{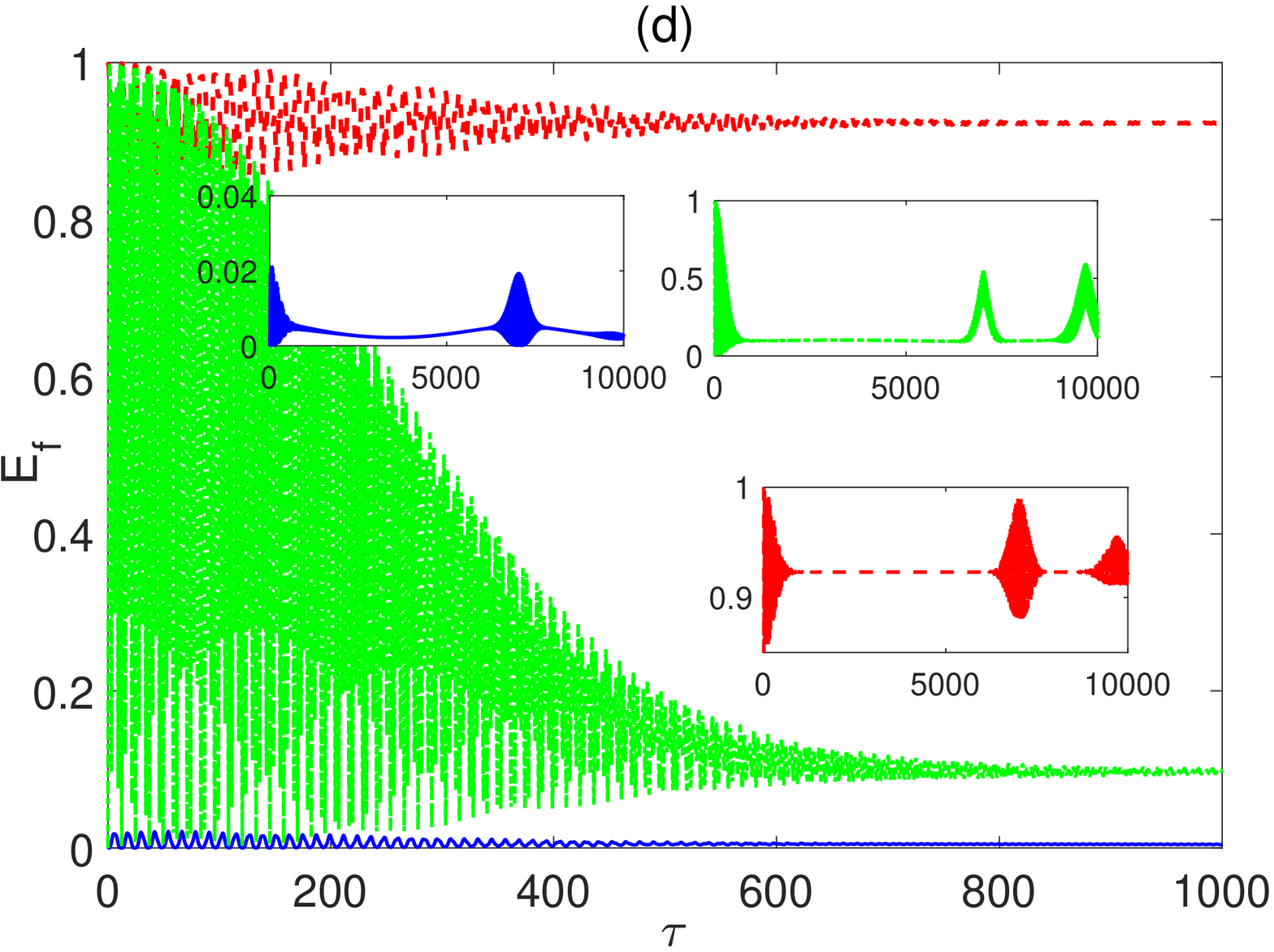}}\\   
  \caption{{\protect\footnotesize (Color online) Time evolution of entanglement versus the scaled time $\tau=\lambda_{2}t$ starting form different initial states at different values of the parameter $\lambda_{1}$ (in units of $\lambda_2$) (a) 1; (b) 0.5 ; (c) 0.1 and (d) 0.01, with detuning $\Delta=0.5$ and $\bar{n}=20$ in all panels. The legend is as shown in panel 8(d).}}
\label{Ent_Lam2_1_Del_05}
 \end{minipage}
\end{figure}
\begin{figure}[htbp]
\begin{minipage}[c]{\textwidth}
 \centering 
\subfigure{\includegraphics[width=8cm]{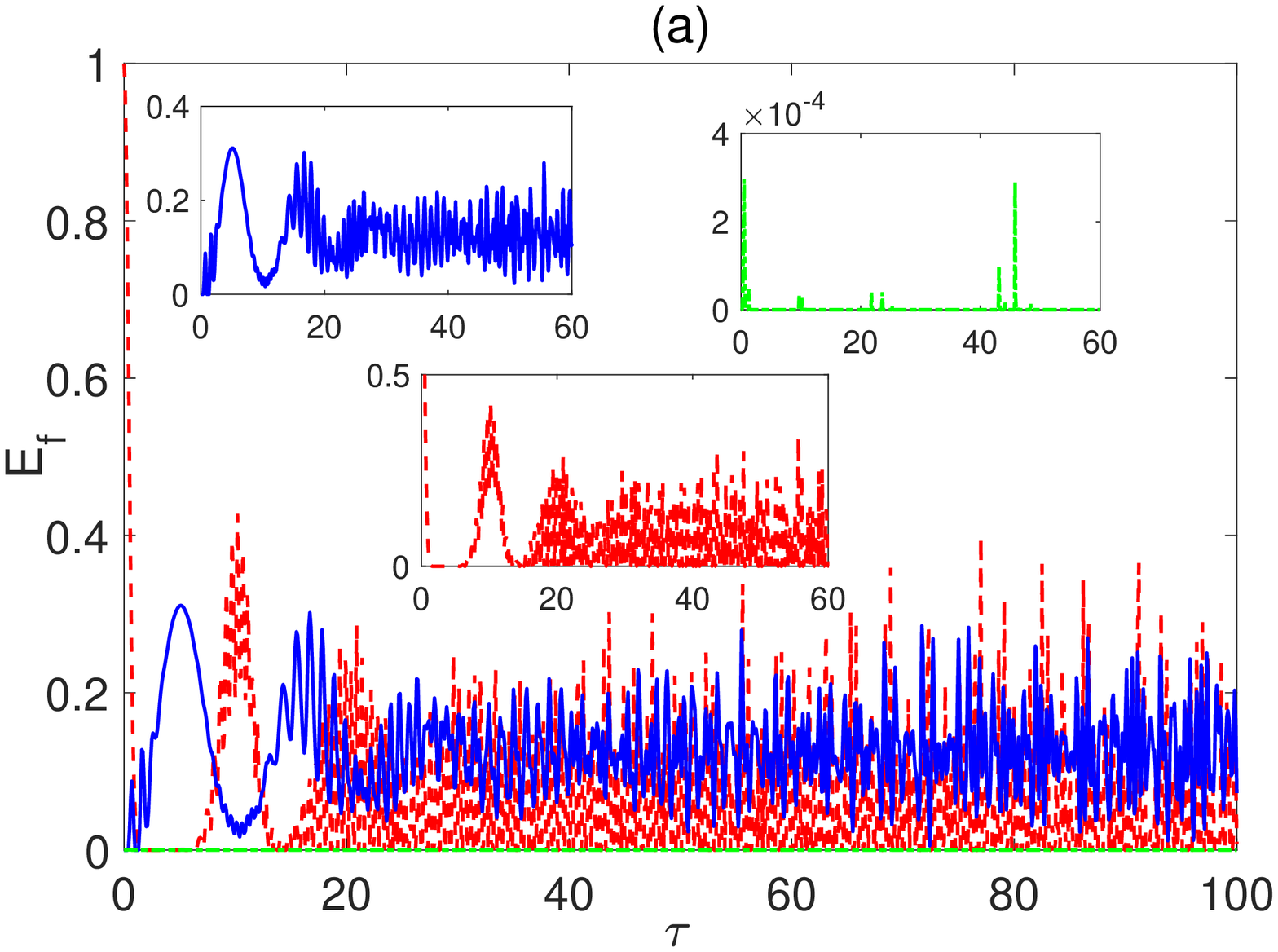}}\quad
   \subfigure{\includegraphics[width=8cm]{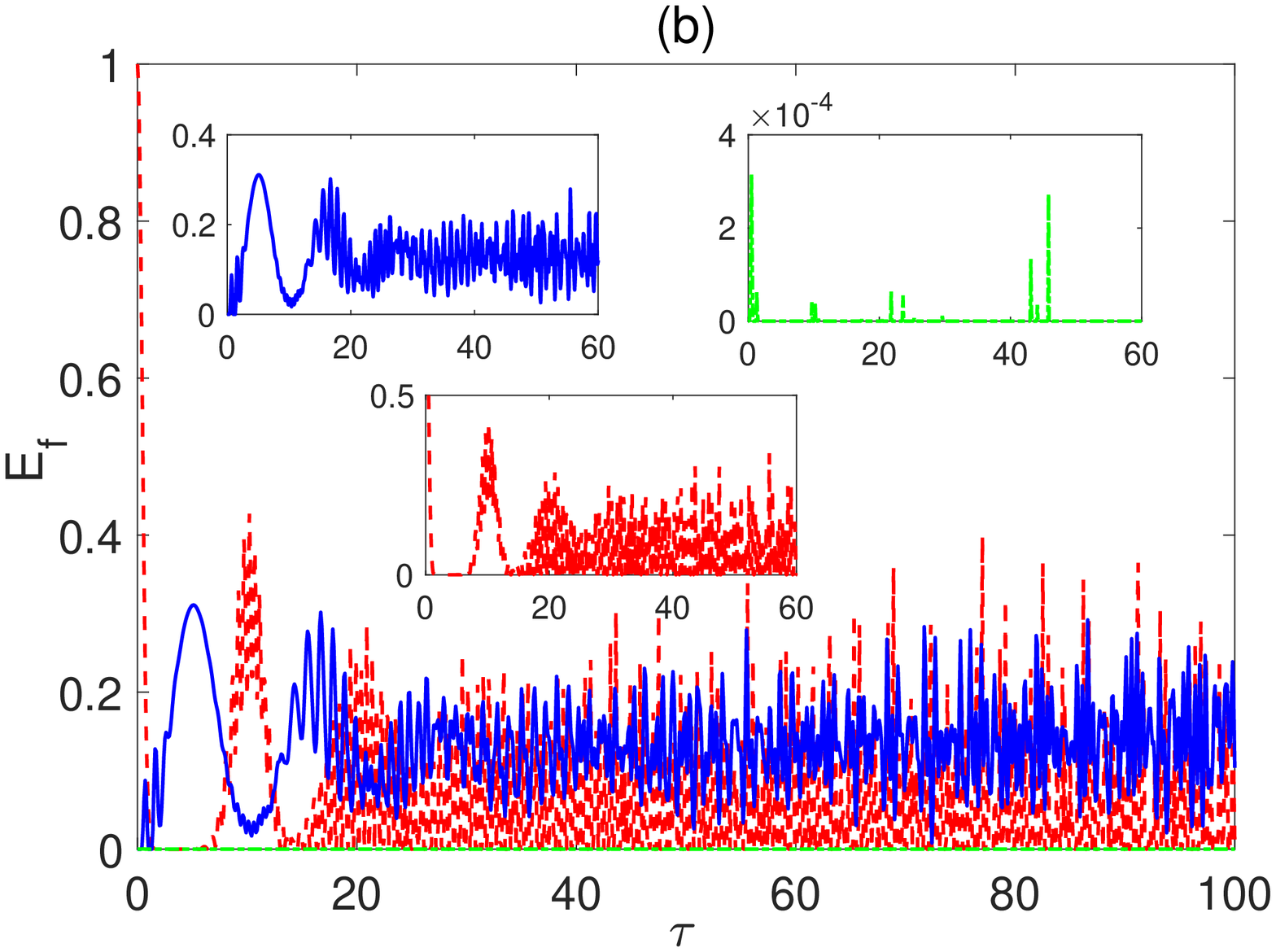}}\\
   \subfigure{\includegraphics[width=8cm]{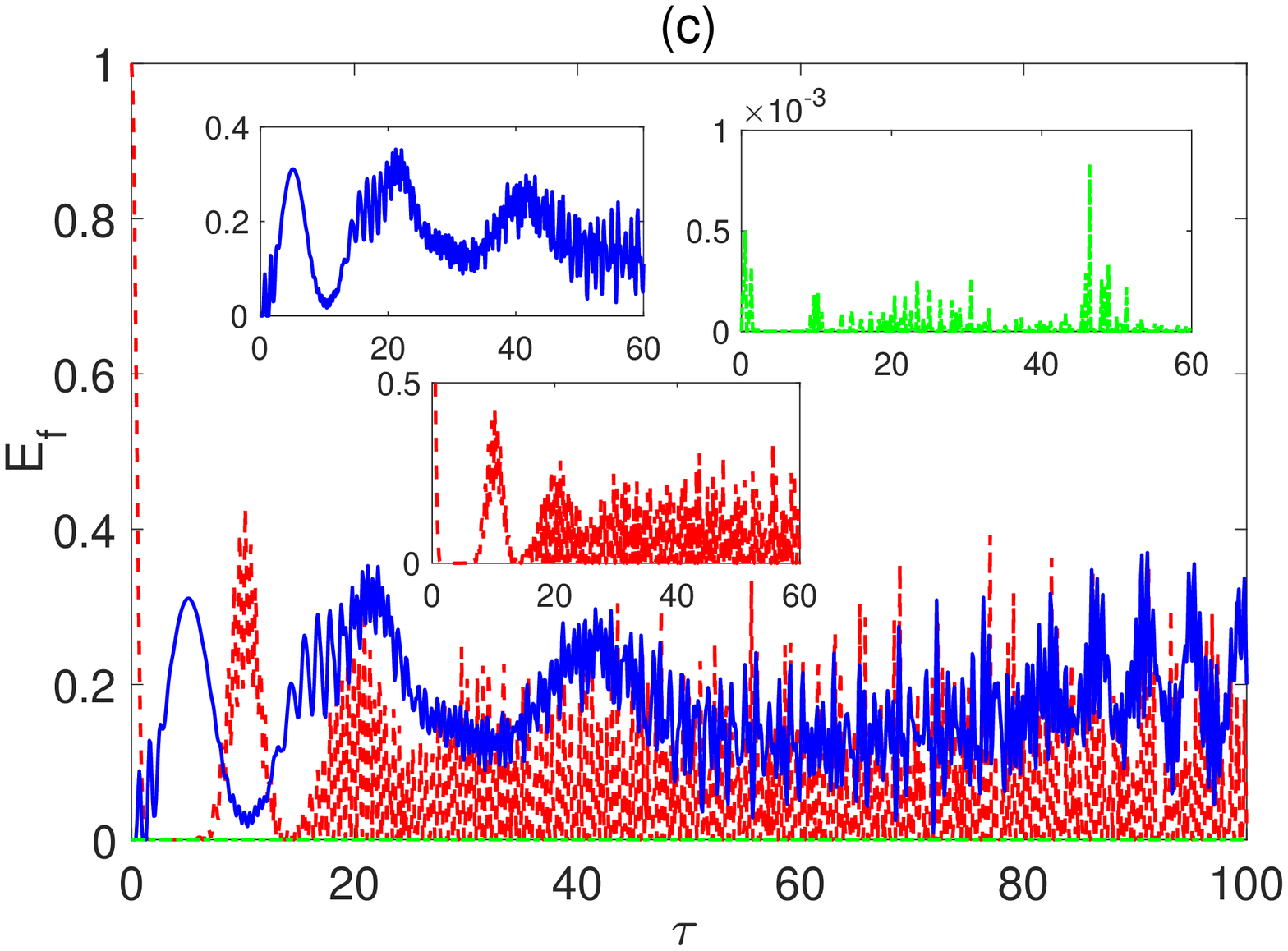}}\quad  
   \subfigure{\includegraphics[width=8cm]{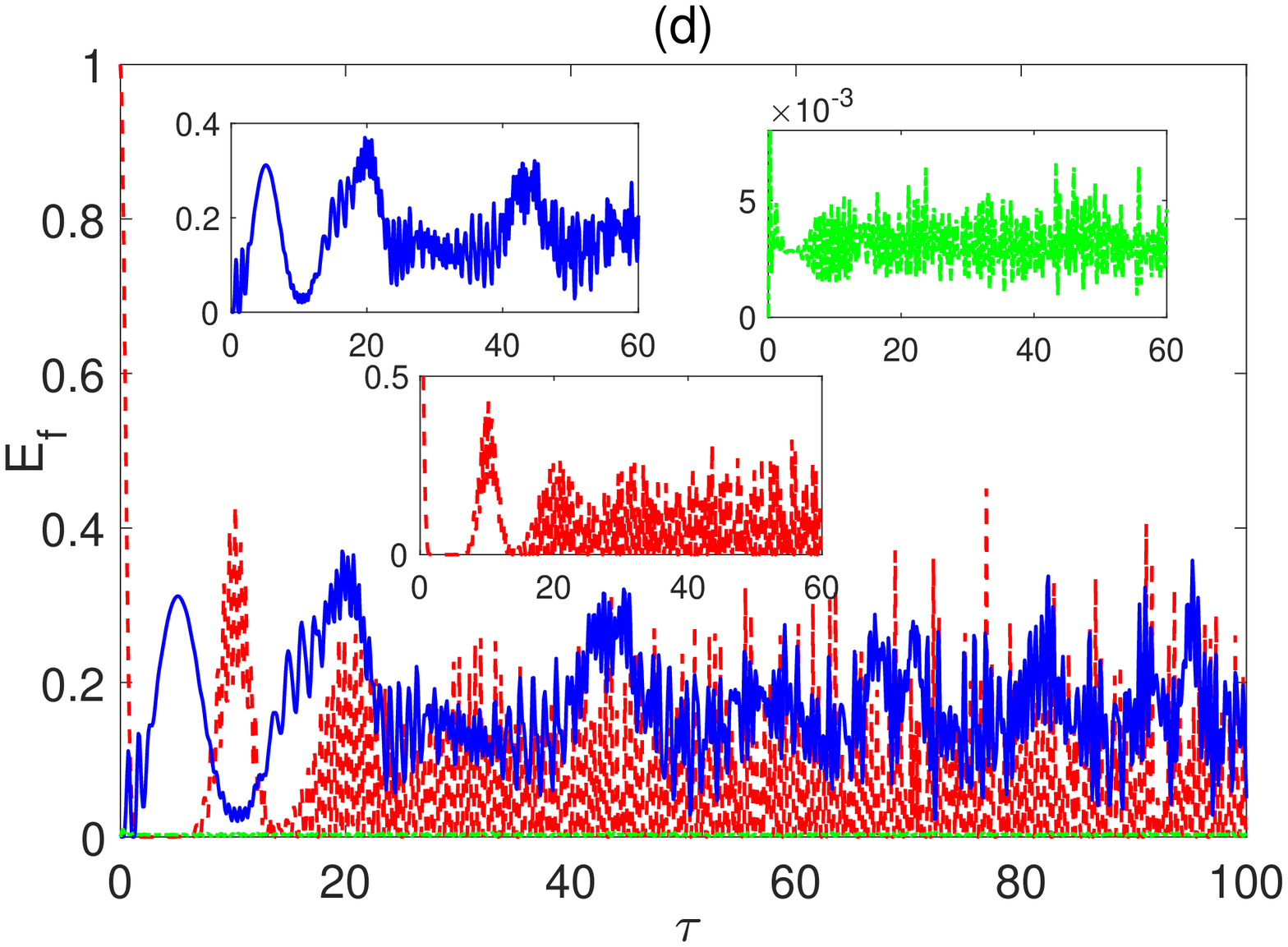}}\\   
  \caption{{\protect\footnotesize (Color online) Time evolution of entanglement versus the scaled time $\tau=\lambda_{1}t$ starting form different initial states at different values of the parameter $\lambda_{2}$ (in units of $\lambda_1$) (a) 0; (b) 0.01 ; (c) 0.1 and (d) 0.5, with detuning $\Delta=0$ and $\bar{n}=10$ in all panels. The legend is as shown in panel 8(d).}}
\label{Ent_n_10_1}
 \end{minipage}
\end{figure}
\begin{figure}[htbp]
\begin{minipage}[c]{\textwidth}
 \centering 
\subfigure{\includegraphics[width=8cm]{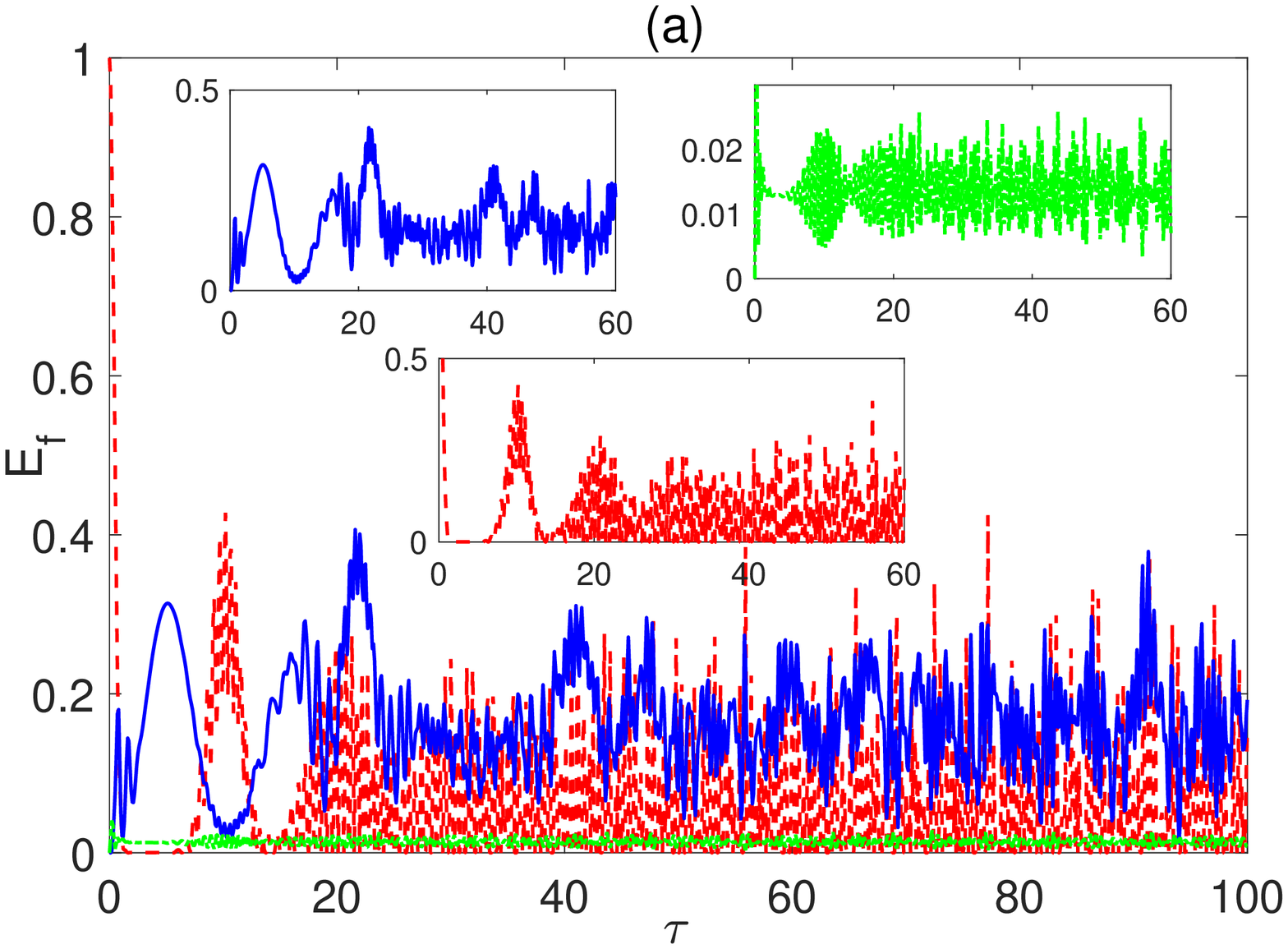}}\quad
   \subfigure{\includegraphics[width=8cm]{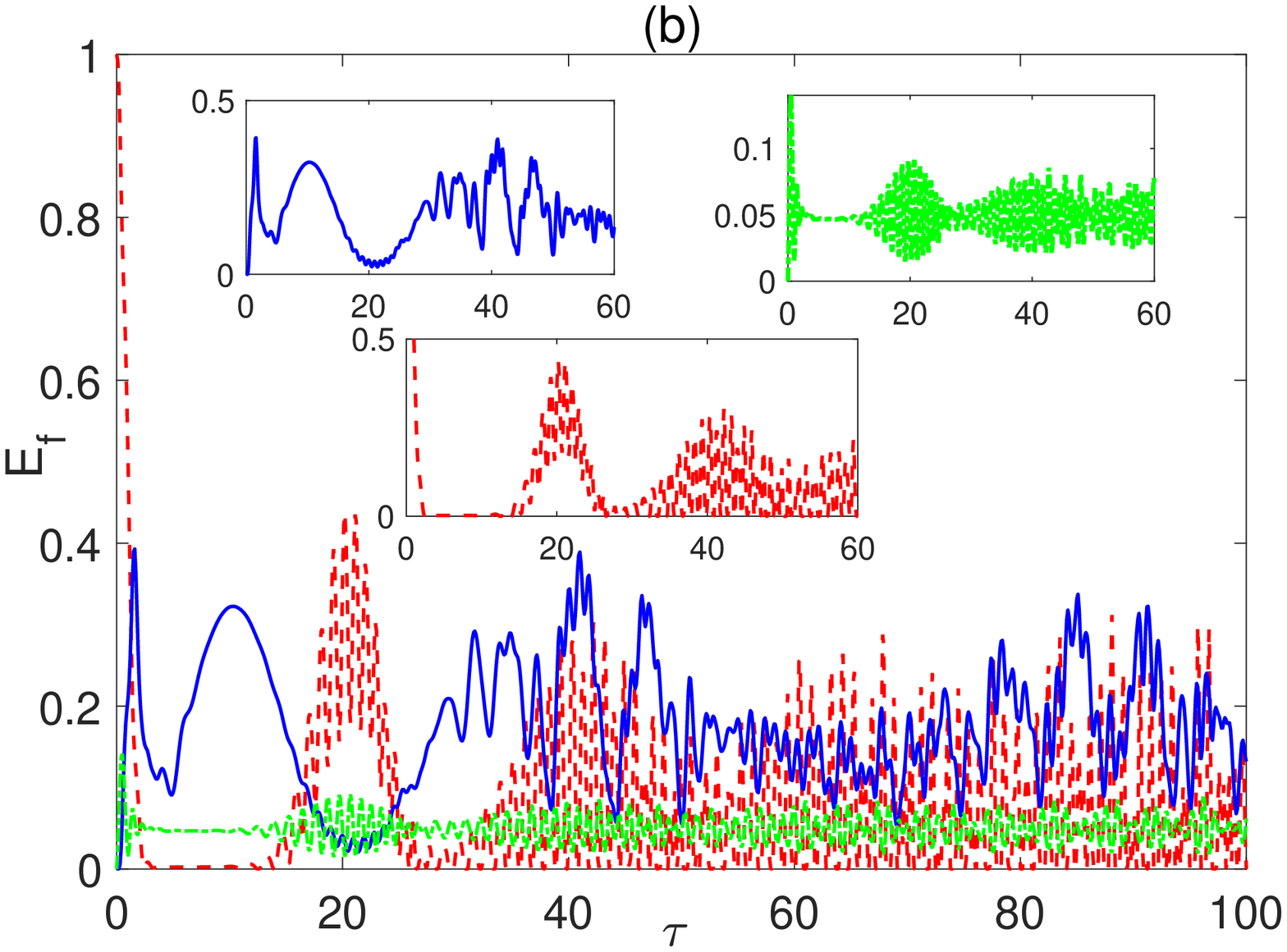}}\\
   \subfigure{\includegraphics[width=8cm]{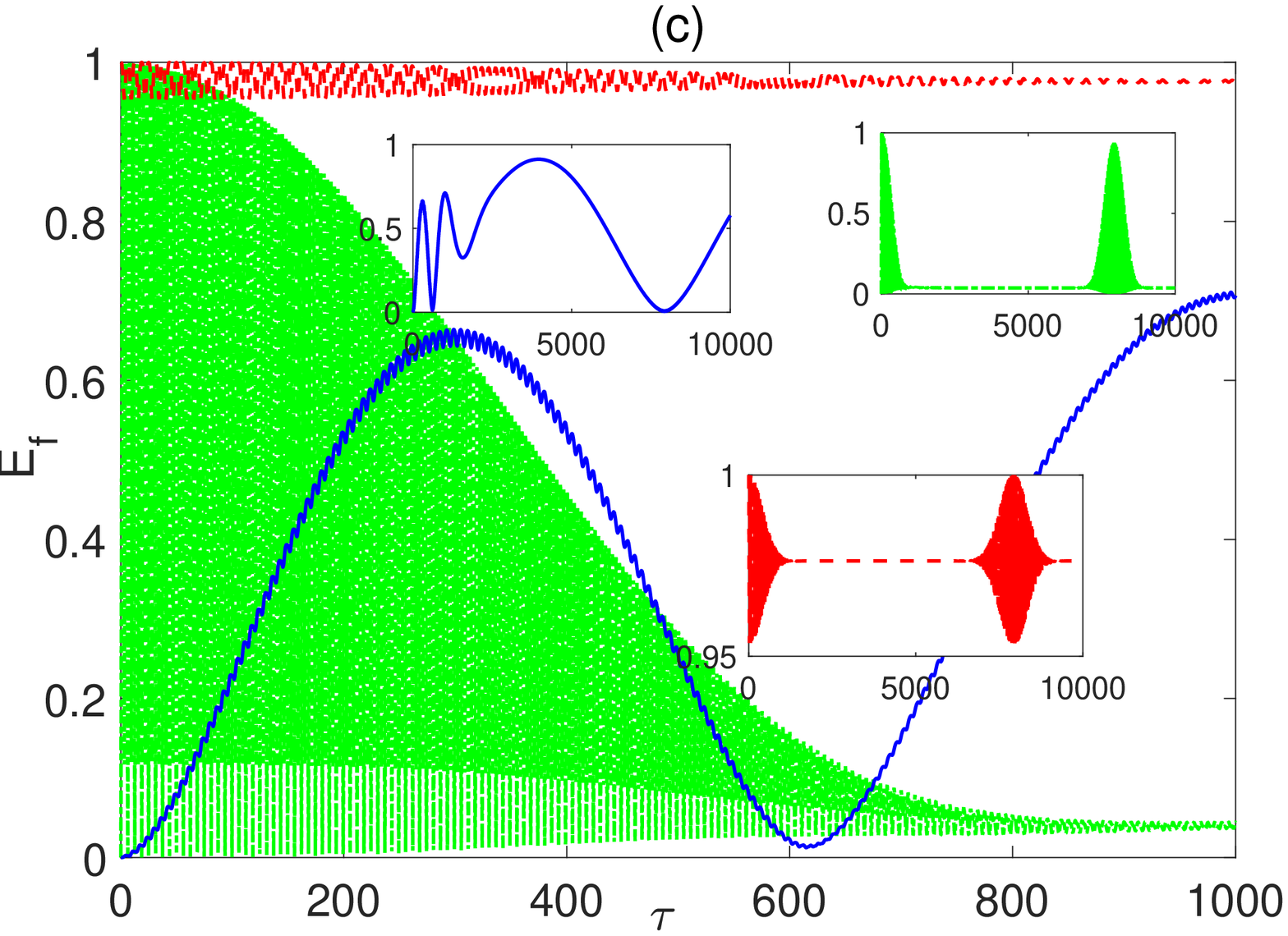}}\\   
  \caption{{\protect\footnotesize (Color online) Time evolution of entanglement versus the scaled time $\tau=\lambda_{2}t$ starting form different initial states at different values of the parameter $\lambda_{1}$ (in units of $\lambda_2$) (a) 1; (b) 0.5 and (c) 0.01, with detuning $\Delta=0$ and $\bar{n}=10$ in all panels. The legend is as shown in panel 8(d).}}
\label{Ent_n_10_2}
 \end{minipage}
\end{figure}

In Fig.~\ref{Ent_Lam1_1_Del_0}, the entanglement $E_{f}$ is plotted versus the scaled time $\tau=\lambda_{1} t$ with the field mean number of photons $\bar{n}=20$, where the two quantum systems are considered starting from three different initial states $\vert\psi_{e}(0)\rangle$, $\vert\psi_{b}(0)\rangle$ and $\vert\psi_{s}(0)\rangle= \frac{1}{2}[\vert e_{1}\rangle \vert e_{2}\rangle+\vert e_{1}\rangle \vert g_{2}\rangle + \vert g_{1}\rangle \vert e_{2}\rangle+\vert g_{1}\rangle \vert g_{2}\rangle]$ at zero detuning. In this figure, we vary the coupling strength $\lambda_2$ (in units of $\lambda_1$) such that $0 \leq \lambda_2 < 1$. It is convenient here to introduce the relative coupling parameter $\lambda_r=\lambda_2/\lambda_1$.
In Fig.~\ref{Ent_Lam1_1_Del_0}(a), we assume no interaction between the two quantum systems whereas each one of them is coupled to the radiation field ($\lambda_r=0$). As can be noticed, the entanglement corresponding to the initial state $\vert\psi_{e}(0)\rangle$ (solid blue line) starts from a zero value then oscillates with a big amplitude and small frequency before eventually turning to a continuous rapid steady oscillation with a smaller amplitude and larger frequency around an average entanglement value of about 0.17. The entanglement evolving from the initial Bell state $\vert\psi_{b}(0)\rangle$ starts with a maximum value before collapsing to zero, then shows a collapse-revival like behavior with a pattern that is not symmetric about the zero value and a revival time $\tau \approx 11$. Starting from the other initial separable state $\vert\psi_{s}(0)\rangle$ (dash dotted green line), the entanglement after oscillating for a quite short period of time, it sustains a zero value with well separated quite small spikes in the order of $10^{-5}$. In Fig.~\ref{Ent_Lam1_1_Del_0}(b), the coupling between the two quantum systems is turned on with a very small value, $\lambda_2=\lambda_r=0.01$, which shows no notable effect on the entanglement dynamics in the three different cases.
As the coupling is increased further to $\lambda_2=\lambda_r=0.1$ and $0.5$, shown in Figs.~\ref{Ent_Lam1_1_Del_0}(c) and (d) respectively, the response of every initial state is different form the others. The entanglement dynamics profile corresponding to the Bell state shows some robustness against the variation of $\lambda_2$ although its revival time decreases to reach $\tau \approx 9$ at $\lambda_r=0.5$ whereas the steady rapid oscillation of entanglement corresponding to the state $\vert\psi_{e}(0)\rangle$ is disturbed and starts to oscillate irregularly with a larger amplitude. The entanglement following form the initial separable state $\vert\psi_{s}\rangle$ reacts differently, where at $\lambda_2=0.1$, it grows up showing an oscillatory behavior with a collapse-revival like pattern, where the entanglement oscillates for a short period of time after $\tau=0$ before collapsing to zero at $\tau \approx 2.5$ and reviving again at $\tau \approx 9$. This behavior becomes even clearer at $\lambda_2=0.5$, as illustrated in the upper right inner panel of Fig.~\ref{Ent_Lam1_1_Del_0}(d), where the oscillation shows a sharper collapse-revival like behavior though it does not collapse to a zero value but rather to a small constant value of about $1.9 \times 10^{-3}$ with a slightly smaller revival time $\tau \approx 8$.

Although the range of the relative coupling $\lambda_r \geq 1$ is not practical, yet we find it interesting to explore and discuss here. In Fig.~\ref{Ent_Lam2_1_Del_0}, we consider that range of values for the coupling parameters, namely $\lambda_2 \geq \lambda_1$ at zero detuning, where the scaled time in this case is $\tau=\lambda_2 t$. We set $\lambda_1$ (in units of $\lambda_2$) as $1$, $0.5$, $0.01$ and $0$ in Figs.~\ref{Ent_Lam2_1_Del_0}(a), (b), (c) and (d) respectively.
In Fig.~\ref{Ent_Lam2_1_Del_0}(a), where $\lambda_2=\lambda_1$ ($\lambda_r=1$), no much of change can be reported compared with Fig.~\ref{Ent_Lam1_1_Del_0}(d) for the entanglement corresponding to the state $\vert\psi_{e}(0)\rangle$. However, the entanglement evolving from the states $\vert\psi_{b}(0)\rangle$ and $\vert\psi_{s}(0)\rangle$ is affected where for the Bell state, the entanglement doesn't collapse to zero any more but to a constant small value of $5.7 \times 10^{-5}$ whereas the constant collapse value of the other state increases to $7.5 \times 10^{-3}$ and the revival time for both states decreases to $\tau \approx 7.5$.
Once we reach $\lambda_r=2$, depicted in Fig.~\ref{Ent_Lam2_1_Del_0}(b), big changes can be noticed in all three cases. The entanglement corresponding to $\vert\psi_{e}(0)\rangle$ oscillates with lower frequency and higher average value. On the other hand, for the other two initial sates $\vert\psi_{b}(0)\rangle$ and $\vert\psi_{s}(0)\rangle$, both of the revival time and collapse constant value of their entanglement start to increase significantly where the revival time becomes $\tau \approx 15$ for both states whereas the constant value turns to $6.62\times 10^{-4}$ and $0.03$ respectively.
When the coupling with the field is enormously reduced, ($\lambda_r=100$), as illustrated in Fig.~\ref{Ent_Lam2_1_Del_0}(c), the amplitude of the entanglement oscillation corresponding to the state $\vert\psi_{e}(0)\rangle$ increases badly reaching $0.79$ with a quite small frequency whereas the entanglement of both of the two other states $\vert\psi_{b}(0)\rangle$ and $\vert\psi_{s}(0)\rangle$ shows a collapse-revival like behavior where it oscillates starting form zero time up to $\tau \approx 650$ before reaching constant values around $0.96$ and $0.07$ respectively then oscillates again at a revival time $\tau \approx 7000$ and keeps repeating the same behavior.
When the coupling between the quantum systems and the radiation field is turned off, presented in Fig.~\ref{Ent_Lam2_1_Del_0}(d), the entanglement shows the expected behavior of the isotropic $XY$ spin-$1/2$ model, where for the initial excited separable state $\vert\psi_{e}(0)\rangle$, it sustains zero value at all times; the other separable state $\vert\psi_{s}(0)\rangle$ evolves to a uniform oscillatory behavior with an amplitude of unity whereas the maximally entangled initial Bell state $\vert\psi_{b}(0)\rangle$ yields maximum entanglement that never change with time.

The impact of a non-zero detuning ($\Delta=0.5$) is explored in Figs.~\ref{Ent_Lam1_1_Del_05} and \ref{Ent_Lam2_1_Del_05}.
Comparing Fig.~\ref{Ent_Lam1_1_Del_05}(a), (b), (c) and (d) with the corresponding ones in Fig.~\ref{Ent_Lam1_1_Del_0}, one can see the main effects of non-zero detuning. The entanglement corresponding to the state $\vert\psi_{e}(0)\rangle$ oscillates with higher amplitude and average value whereas the entanglement of the state $\vert\psi_{b}(0)\rangle$ takes slightly smaller revival time in every panel in Fig.~\ref{Ent_Lam1_1_Del_05} compared with Fig.~\ref{Ent_Lam1_1_Del_0}. For the state $\vert\psi_{s}(0)\rangle$, the non-zero detuning reduces both of the average value of the entanglement oscillation and its revival time. 
Now comparing Fig.~\ref{Ent_Lam2_1_Del_05} panels with the corresponding ones in Fig.~\ref{Ent_Lam2_1_Del_0}, one can see that the entanglement corresponding to $\vert\psi_{e}(0)\rangle$ shows a collapse-revival like profile for the first time at $\lambda_r=10$ and $100$. as shown in Figs.~\ref{Ent_Lam2_1_Del_0} (c) and (d) although in the collapse periods the entanglement doesn't take either a zero or even a constant value but shows very small variations in magnitude compared with the revival periods. 
The entanglements corresponding to the other two states suffer reduction in both of the revival time and the collapse constant value in all cases of Fig.~\ref{Ent_Lam2_1_Del_05} compared with that of Fig.~\ref{Ent_Lam2_1_Del_0}. As $\lambda_r$ increases, the revival times corresponding to the three different cases become very close and reach $\tau \approx 6400$ at $\lambda_r=100$. 

In Figs.~\ref{Ent_n_10_1} and \ref{Ent_n_10_2} we test the effect of varying the field intensity by reducing the mean number of photons to $\bar{n}=10$ at zero detuning. Comparing these two figures with the corresponding figures~\ref{Ent_Lam1_1_Del_0} and \ref{Ent_Lam2_1_Del_0} where $\bar{n}=20$, one can notice the impact of decreasing the field intensity. In general, the entanglement oscillation becomes more rapid with higher frequency and larger amplitude and average value. The collapse-revival like pattern of the entanglement is not as sharp as it was in the case of $\bar{n}=20$ and the revival time is reduced in agreement with what we have observed in the population inversion study. In other words, increasing the field intensity enhances the collapse-revival like behavior of the entanglement and reduces the average value of the entanglement oscillation, which increases as the coupling $\lambda_2$ increases.   

\section{Conclusion}
We studied the time evolution of a composite system consisting of two interacting identical two-level quantum systems modeled as two coupled spin half particles with Heisenberg isotropic XY exchange interaction. Each one of the particles is coupled to the same single mode radiation field with the same coupling strength, where the field was considered to be in a coherent state. This composite system could be realized in cavity (circuit) QED. We presented an exact analytic solution of the problem which spans the entire parameter space of the system. The effect of introducing and varying the interaction between the two quantum systems in presence of the radiation field on the dynamical properties of the system was investigated. Starting with a disentangled state where the two quantum system are in the upper levels (excited state), the famous collapse-revival behavior of the population inversion of any of the two quantum systems was found to vary depending on the ratio between the two couplings (system-system to system-radiation) $\lambda_r$. As this ratio was increased gradually until reaching one, the revival pattern split into smaller patterns, which indicates that a strong coupling between the two quantum systems may interrupt the continuous energy exchange between the quantum systems and the radiation field within the revival intervals. The system-system coupling showed a similar impact when a non-zero detuning was introduced. Applying a radiation field with higher intensity caused a longer revival time, as usual, but furthermore it reduced the effect of the system-system coupling, where higher coupling strength was needed to split the revival pattern. Testing the system starting from a different initial state, a maximally entangled Bell state, the population inversion showed a different dynamics at zero detuning, where sharp continuous oscillation was observed but when a non-zero detuning was applied the population inversion behavior showed a collapse-revival like pattern.

The bipartite entanglement between the two quantum systems was investigated as well starting from three different initial states. The first was a disentangled state with the two quantum systems in the upper excited states, the second was a disentangled state with all different combinations of the two quantum systems states and the third was a maximally entangled Bell state. The time evolution of the entanglement was found to vary significantly depending on the coupling ratio $\lambda_r$.
When the two quantum systems were decoupled from each other but interacted with the radiation field, the bipartite entanglement dynamics varied depending on the initial state of the composite system, where it showed either an oscillatory behavior, a collapse-revival like pattern or maintained a zero value with separated spikes. Turning on the interaction between the two quantum systems caused mainly an enhancement of the entanglement, which raised its oscillation average value particularly for the initial disentangled states cases whereas the Bell state entanglement showed some robustness for as long as the coupling ratio $\lambda_r < 1 $. Enhancing the entanglement by increasing the two systems mutual interaction made the collapse-revival like profiles, which is a sign of energy exchange between the radiation field and each one of the two quantum systems, more visible. Nevertheless as the interaction between the two quantum systems was increased further such that $\lambda_r \geq 1$, the exchange process was delayed where the revival time increased significantly. Increasing the field intensity caused mainly a sharper collapse-revival like behavior of the entanglement but reduced the average of the entanglement oscillation. The non-zero detuning in the system caused different effects based on the initial state particularly at $\lambda_r<1$ but overall it enhanced the collapse-revival like behavior of the entanglement particularly at large $\lambda_r$ value and reduced its the revival time.

It is interesting, in the future, to consider two non-identical quantum systems with anisotropic mutual interaction between them and different coupling strength with the radiation field to find out the effect of such asymmetry on the dynamical behavior of the system compared with the current model.

\end{document}